\documentclass[useAMS,usenatbib,usegraphicx,preprint]{mn2e}
\usepackage{graphicx}
\usepackage{setspace}
\usepackage{epsfig}
\usepackage{ulem}

\newcommand{\kms}{\>{\rm km}\,{\rm s}^{-1}}
\newcommand{\masyr}{\>{\rm mas}\,{\rm yr}^{-1}}

\newcommand{\Msun}{\>{M_{\rm \odot}}}

\newcommand{\poubelle}[1]{}

\def\aj{AJ}%
\def\araa{ARA\&A}%
\def\apj{ApJ}%
\def\apjl{ApJ}%
\def\apjs{ApJS}%
\def\apss{Ap\&SS}%
\def\aap{A\&A}%
\def\aaps{A\&AS}%
\def\mnras{MNRAS}%
\def\na{New A}%
\def\pasa{PASA}%
\def\nat{Nature}%
\def\pmora #1{snapshot_#1_im.ps}

\def\mldyn {$(M/L)_{\rm dyn}$}

\title{Reproducing properties of MW dSphs as descendants of DM-free TDGs}
\author[Yanbin Yang et al. ]{Yanbin Yang$^{1,2}$\thanks{E-mail: yanbin.yang@obspm.fr}, Fran\c cois Hammer$^{1}$\thanks{E-mail: francois.hammer@obspm.fr}, Sylvain Fouquet$^{1}$, Hector Flores$^{1}$, \newauthor Mathieu Puech$^{1}$, Marcel S. Pawlowski$^{3,4}$,   Pavel Kroupa$^{3}$\\
$^{1}$Laboratoire GEPI, Observatoire de Paris, CNRS-UMR8111, Univ Paris Diderot, 5 place Jules Janssen, 92195 Meudon France\\
$^{2}$National Astronomical Observatoires, Chinese Academy of Sciences, 20A Datun Road, Chaoyang District, Beijing 100012, China\\
$^{3}$Helmholtz-Institut fuer Strahlen und Kernphysik (HISKP) Nussallee 14-16, 53115 Bonn, Germany\\
$^{4}$Department of Astronomy, Case Western Reserve University, 10900 Euclid Avenue, Cleveland, OH 44106, USA}
\begin{document}

\date{Accepted; Received; in original form}

\pagerange{\pageref{firstpage}--\pageref{lastpage}} \pubyear{?}

\maketitle

\label{firstpage}

\begin{abstract}
The Milky Way (MW) dwarf spheroidal (dSph) satellites are known to be the most dark-matter (DM) dominated galaxies with estimates of dark to baryonic matter reaching even above one hundred. It comes from the assumption that dwarfs are dynamically supported by their observed velocity dispersions. However their spatial distributions around the MW is not at random and this could challenge their origin, previously assumed to be residues of primordial galaxies accreted by the MW potential. Here we show that alternatively, dSphs could be the residue of tidal dwarf galaxies (TDGs), which would have interacted with the Galactic hot gaseous halo and disk. TDGs are gas-rich and have been formed in a tidal tail produced during an ancient merger event at the M31 location, and expelled towards the MW. Our simulations show that low-mass TDGs are fragile to an interaction with the MW disk and halo hot gas. During the interaction, their stellar content is progressively driven out of equilibrium and strongly expands, leading to low surface brightness feature and mimicking high dynamical M/L ratios. Our modeling can reproduce the properties, including the kinematics, of classical MW dwarfs within the mass range of the Magellanic Clouds to Draco. An ancient gas-rich merger at the M31 location could then challenge the currently assumed high content of dark matter in dwarf galaxies. We propose a simple observational test with the coming GAIA mission, to follow their expected stellar expansion, which should not be observed within the current theoretical framework.
\end{abstract}



\setcounter{figure}{0}

\section{Introduction}
So far, 26 dwarf galaxies have been discovered within 
250~kpc from the MW \citep{2012AJ....144....4M}. 
Among them 11 bright ``classical'' dwarf galaxies, i.e., 
9 dSphs plus the two Magellanic clouds (MCs) 
have been studied intensively
\citep[see e.g.,][]{1998ARA&A..36..435M,2012AJ....144....4M}. 
Possibly the most spectacular property of dSphs is that together with ultra-faint dwarfs, they appear to be dark matter
dominated galaxies \citep{1998ARA&A..36..435M,2009ApJ...704.1274W}.
Large amounts of dark matter were firstly inferred from the high value of
the dynamical mass-to-light ratio, ($M/L$)$_{\rm dyn}=31$, for Draco by \citet{1983ApJ...266L..11A}. 
This result was  confirmed by \citet{1983ApJ...266L..17F} for most MW dSphs. 
Stellar kinematics based on large samples of stars have confirmed with an improved precision the high ($M/L$)$_{\rm dyn}$ values for MW dSphs as shown by \cite{2005nfcd.conf...60E} 
and \citet{2009ApJ...704.1274W}.
Assuming an isotropic spherical system, the dynamical mass-to-light ratio was firstly estimated using 
the King formula \citep[e.g.,][]{Rood1972,1986AJ.....92...72R,1995AJ....109.1071P}:
\begin{equation}
\label{eqml}
\big(\frac{M}{L}\big)_{\rm dyn} \approx \frac{9}{2\pi G} \frac{\sigma_0^2}{\mu_0r_{\rm c}},
\end{equation}
where $G$ is the gravitational
constant, $\sigma_0$ the central line-of-sight velocity dispersion,
$\mu_0$ the central surface brightness,
and $r_{\rm c}$ the core radius defined at half the central surface brightness. 
In more recent works most authors use the total dynamical mass-to-light ratios \citep[e.g.,][]{1998ARA&A..36..435M,2009ApJ...704.1274W}. As the two different methods provide us systematically consistent estimations of the dynamical mass-to-light ratios, we adopt Eq. (1) in the following.
The presence of large amounts of dark matter in dSphs is also supported 
by the flatness of velocity dispersion profiles \citep{2007ApJ...667L..53W,2009ApJ...704.1274W},
implying that the mass-follows-light models under-predict the observed velocity dispersions at large radii.  
This is the strongest evidence that dSphs have dominant and extended dark matter halos.

These studies of dark matter content all assume that
MW dSphs are in virial equilibrium \citep[][and references therein]{2012arXiv1205.0311W}.
Tidal forces may have played an important role 
in reshaping the nearby MW companions \citep{1997NewA....2..139K}. A robust example is the Sagittarius dwarf
spheroidal \citep{1994Natur.370..194I}. 
Its tidal debris has been discovered on its orbit around the MW 
\citep{2003ApJ...599.1082M}.  
Recent observations also suggest that some of the classical dSphs possess
an extended stellar profile beyond their tidal radius,
such as 
Carina \citep{2005AJ....130.2677M,2006ApJ...649..201M,2012arXiv1211.4875B},
UMi, Draco \citep{Palma2003,2005ApJ...631L.137M},
Leo~I \citep{2007ApJ...663..960S},
Sextans \citep{1995MNRAS.277.1354I}.
Using high resolution simulations including ram pressure and tidal stripping,
\citet{2006MNRAS.369.1021M} proposed that dSphs may be 
the descendants of gas-rich dwarf irregular galaxies 
that have lost their gas due to the interaction 
with the hot MW corona, as suggested from observations \citep{Blitz2000,Grebel2003}. 
This evolutionary scenario has shown an impressive success in reproducing dSph properties. It assumes that the progenitors of MW
satellites have a cosmological origin, i.e., they are relics of primordial dwarfs and as such, they do contain dark matter.

Perhaps the geometrical distribution of the dSphs surrounding both the MW and M31 is as surprising as their heavily dominant dark matter content.
\citet{1983IAUS..100...89L} already noticed that the then known satellite galaxies and some globular clusters form a strip on the sky and suggested a common origin for them. \citet{2005A&A...431..517K} investigated the distribution of satellite galaxies and showed them to lie along in a disk of satellites (DoS).
\cite{2012MNRAS.423.1109P}  found that the classical MW dwarfs, ultra-faint satellites, streams and young halo globular clusters, are likely settled in a Vast POlar
Structure (VPOS) almost perpendicular to the MW disk sharing the same angular momentum as the DoS. Associating it with streams suggests a tidal origin for the VPOS.
\citet{2008ApJ...680..287M} found that 6 \citep[and presently 9, see e.g.,][]{Fouquet2012} of the 11 classical dSphs have 
similar angular momentum directions indicative of their coherent motions within the VPOS.
It has been suggested \citep{2005A&A...431..517K,2007MNRAS.374.1125M,2008ApJ...680..287M,2010A&A...523A..32K,Pawlowski2012b,Pawlowski2013b} that the VPOS might not be fully consistent with cosmological simulations \citep{2005ApJ...629..219Z,2005MNRAS.363..146L,2009MNRAS.399..550L,2011MNRAS.415.2607D}.

The new discovery of a Vast Thin Disk of Satellites (VTDS) around M31 by \citet{Ibata2013} appears even more puzzling. Half of the M31 dSphs are within a 400 kpc co-rotating plane whose thickness of 14 kpc cannot be reproduced by infall from large filaments \citep{Ibata2013}.  The significance of this feature is high even if one accounts for the clustering of dSphs \citep{Fattahi2013}, and
a tidal origin is also suggested (Hammer et al.\ 2013, but see Shaya \& Tully 2013 for another viewpoint).
It is now acknowledged that gas-rich mergers may produce spirals galaxies \citep{Brook2004,Hopkins2009,Hammer2005,Hammer2009} that may significantly contribute to the growth of types earlier than Sc \citep{Puech2012}. M31 is a Sb galaxy which is a good candidate to have experienced a merger 6 billion years ago \citep{Hammer2010}, given its  bulge Sersic index n=2.2 \citep{Courteau11} and the metal-rich stellar content in its halo, including the Giant Stream. \citet{Fouquet2012} suggested that most of the MW satellites may originate as TDGs from a tidal tail expelled from this event \citep[see also][]{Hammer2010,2010ApJ...725L..24Y}. TDGs are distributed along tidal tails and their motions are more appropriate to reproduce the above mentioned vast structures \citep[see e.g., the reproduction of the MW dSph plane by][]{Fouquet2012}. As the VPOS, the VTDS is almost perpendicular to the MW disk and moreover it points towards the MW within only 1 degree. This has lead us to propose a quite radical scenario, for which both relic structures are related to a single, ancient event in the Local Group \citep{Hammer2013}. Both the VPOS and the VTDS are reproduced by the same family of models, though improvement is desirable in a context of an expected huge number of parameters and constraints. Although recently \citet{Bahl2014} claim that structures like the VTDS are common in cosmological simulations, their methodology may lead to a biased conclusion. Their claims have been rebutted by \citet{Ibata2014} and \citet{Pawlowski2014}.

The above proposition (hereafter M31 scenario) implies that the M31 merger may contribute to the formation of  large-scale structure of galaxies in the Local group and that gas-rich TDGs may survive in the Local Group if they are far enough from the M31 and MW, escaping from the ram-pressure stripping of hot gaseous halos.
In particular, the gas-rich dwarf irregular IC~10  lies in the plane of VTDS and is close to the first tidal tail of the M31 merger \citep[see Figure 3 in][]{Hammer2013}. The latest estimation of its mass-to-light ratio is 2 \citep{Tollerud2014}, which favors the idea that IC 10 could be a TDG candidate as a relics of the M31 merger. 
In order to understand the impact of the M31 merger to the formation of the structure of galaxies in the Local Group, such as the great planes discovered by \citet{Pawlowski2013a} and \citet{Pawlowski2014}, much more thorough studies have to be performed.

The  M31 scenario implies that the MW dSphs would be the descendants of TDGs that were ejected 8.5 Gyr ago
from the M31 merger. Initially their stars were
extracted from the M31 progenitors, so they contain
aged stars similar to those observed in many dSph.
In the M31 scenario, TDGs are expected to evolve
for several billion years in a tidal tail, before being
accreted by the MW. Their stellar content is then the
sum of an aged stellar population extracted from the
progenitors with intermediate aged stars formed during
their evolution within a gas-rich medium of the tidal tail.
In their statistical observational study of TDGs, \citet{Kaviraj12} conclude
that on average, half of their stellar population is made by stars
drawn from the parent progenitor. 

It would be important to follow the star formation
in TDGs during such a large period, which requires
a full detailed study that is out of the scope of this paper.
However it could be expected that massive TDGs are likely to accrete
more gas from the tidal tail and then to have a more active star formation
history than their lower-mass counterparts. This is also expected from
the Schmidt-Kennicutt law, i.e., low mass
TDGs possibly do not reach the critical gas density to
form many new stars. 
Some MW dSphs such as  UMi, Draco and Sextans hold oldest populations \citep{Dolphin2005}.
UMi, as well as Sextans does not present intermediate age \citep[see][]{Cohen2010,Lee2009} conversely to Draco \citep{Cohen2009}, though the old population
is from 8-9 to 13.5 Gyrs, with uncertainties allowing some star formation down to 5-6 Gyrs.
In other words, star formation histories of TDGs
are not different than those of other galaxies, except by the fact that they have
to live in a gas-rich environment, letting Hammer et al. (2013) to
assume that it could be generally consistent with the observed stellar mass-metallicity relation of dwarf galaxies in the Local Group \citep{Kirby13}. 
\citet{Recchi2014} suggest that the chemical abundance of TDGs may fit the mass-metallicity relation of dwarf galaxies in the Local Group if they were formed at early epochs, e.g., $\sim$~10 Gyr ago, from less enriched gas.
They use standard models of chemical evolution \citep{Recchi2008} with outflow rates larger in smaller galaxies \citep[as in][]{Spitoni2010} but applying the IGIMF (the galaxy-wide, SFR-dependent IMF). 
The same models with higher degrees of pre-enrichment reproduce well the oxygen content of younger TDGs
such as those observed by \citet{Duc2014} and \citet{Boquien2010}.

The M31 scenario implies that the MW dSph progenitors are TDGs \citep[i.e., galaxies devoid of dark matter,][]{Barnes1992,Hibbard2001}, as formerly suggested by \citet{1989ApJ...341L..41K,1993ApJ...409L..13K,1995ApJ...442..142O,Dabringhausen2013} and modelled by \citet{1997NewA....2..139K} as confirmed by \citet{1998ApJ...498..143K}, \citet{Metz2007} and  \citet{2012MNRAS.424.1941C}. However these models need several orbits for the gas-deficient TDGs to be transformed into dSphs. The same applies to the \citet{2006MNRAS.369.1021M} model of gas-rich, dark-matter dominated disk galaxies being progenitors of dSphs. It may become problematic since dSphs share the same orbit as the Magellanic Clouds which have a large proper motion \citep[see e.g.,][]{Vieira2010}, and their large amount of gas suggest them to be at first approach. 
In this paper we propose to follow simulated TDGs selected from the first tidal tail that is created 8.5 Gyr ago by the first passage of a merger in M31. The initial TDG velocities are from the orbital motion of the LMC using the most recent results compiled
by \citet{Vieira2010} with quite large uncertainties (see their fig.\ 10). It corresponds to the values of the model developed by  \citet{Fouquet2012} while the amplitude of the LMC velocity would have been too large if using the Kallivayalil et al. (2009) value.
This scenario implies that the MW satellites were on unbound orbits 
when they entered the MW for the first time. 
This is consistent with independent studies which support
a first infall scenario of MW satellites, 
such as \citet{2007ApJ...668..949B} for the LMC, 
\citet{2011AJ....142...93M} for Fornax
and \citet{Sohn2013} for Leo I.

In this paper, we carry out N-body/SPH simulations of the interactions
between gas-rich TDGs and the MW in order to test whether gas-rich TDGs could be 
the progenitors of MW dSphs in the M31 ancient merger scenario.
In Sect.~2 we describe the configurations of the simulations, including 
initial conditions (ICs) of the MW and the TDGs and the orbit setups.
In Sect.~3, we present the results of the simulations.
In Sect.~4, we propose a mechanism that could explain the MW dSphs.
In Sect.~5, we compare our results to the observations.
In Sect.~6, we discuss the nature of the MW dSphs in view of our simulations.
Conclusions will be drawn in Sect.7.

\section{Simulations}
We used Gadget2 \citep{2005MNRAS.364.1105S} with gas
cooling, feedback and star-formation as developed by
\citet{Wang2012}. Here, we describe the initial conditions
and orbit configurations.

\subsection{Initial conditions of the MW}
\begin{figure}
\includegraphics[width=8cm]{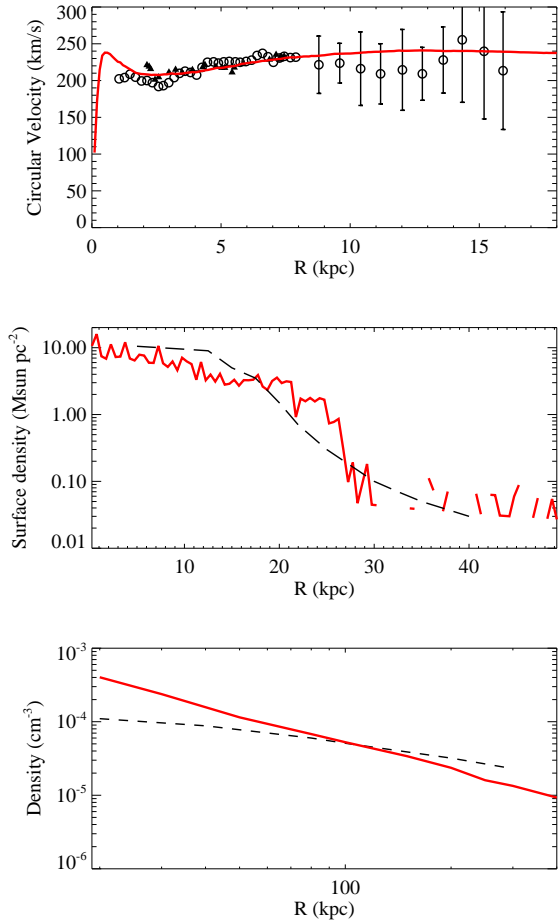}
\caption{Mean properties of the MW model. {\it Top panel:} Rotation curve, which is derived from the total mass enclosed within a given radius,
  of our MW model (red line),  with comparison to
  the observed rotation curve of the MW which is adapted from \citet{2002ApJ...573..597K}: 
  open circles for the data from \citet{1985AJ.....90..254K} and filled
  triangles for those from \citet{1986A&AS...66..373K}. 
  {\it Middle panel:} Surface density profile of the gas disk of the MW model (red line). 
  The observed H\,{\sc i} gas distribution of the MW disk \citep{Kalberla09} is shown as a long-dashed line.
  {\it Bottom panel:} Density of hot gas in the MW model (red solide line) in comparison
  to the core model (dashed line) described in \citet{Fang2013}.
}
\label{figmwrc}
\end{figure}

We have constructed a MW model with four components: a dark matter halo, a gaseous halo,
a bulge and a disk (star+gas). The N-body realization of the MW model has been created following 
\citet{1988ApJ...331..699B,2002MNRAS.333..481B} using {\sc Zeno}\footnote{
  http://www.ifa.hawaii.edu/$\sim$barnes/software.html} \citep[e.g.,][]{1992ARA&A..30..705B}.
As investigated by \citet{2010AdAst2010E..25M},
ram-pressure stripping may play a very important role during 
interactions between gas-rich dwarf irregulars and the MW. 
The former may loss completely their gas 
in 2 Gyr and then be transformed into dwarf spheroidals consequently. 
The TDGs that are formed in the M31 model are gas-rich. 
Thus a gaseous halo in the MW model is mandatory to be included
in our simulations in order to better understand
the interaction between gas-rich TDGs and the MW.

Following \citet{Mastropietro2005} and \citet{Mayer2001a}
we have first set up a hot gaseous halo in hydrostatic equilibrium with a mass of $10^{10}\Msun$, 
which follows the density distribution of the dark halo
and a mean temperature of $5.5 \times 10^5$~K. 
This model provides a mean density of hot gas of 6.4, 0.53, and 0.05 $\times 10^{-5}\,\rm{cm}^{-3}$ at
50, 100, and 300 kpc, respectively. This density distribution is not fully consistent with 
observations at large radii. 
For example, \citet{Sembach2003} suggest
that gas in the MW halo at distances larger than 70 kpc may have a density of $10^{-5}$-$10^{-4}$~cm$^{-3}$
and that this density of gas may extend into the Local Group medium.
With this MW model, we did a large number of simulations of the TDG-MW interaction
following the orbits described in Sect.~\ref{secorb}. 
We found a set of models which could reproduce well
the physical properties of some dSphs, such as Draco and UMi, including their high M/L. 
In this series of models, transforming a gas-rich TDG into a dSph 
requires small pericenters, i.e., $<25$~kpc.
However, when comparing these results to the pericenters of the MW dSphs (Sect.\ref{secfit}), 
we found that most MW dSphs have large pericenters, typically $>40$ kpc.
The lack of gas removal on large pericentric orbits may be simply due to the low density of hot gas at large radii.

More recently, X-ray observations have further confirmed
that the MW is indeed surrounded by a hot gaseous halo 
with temperature of $\approx\!10^6$~K out to large radii. 
With a limited data-set of absorption features of the MW hot gas 
towards QSOs, models of the MW hot halo
have been constructed \citep{Fang2013}. 
This suggests that the total mass of the MW hot gas may reach $5\times 10^{10}\Msun$ within the virial radius, $R\approx 260$~kpc. 
Taking this model of hot halo, we constructed another galaxy model which 
gathers the essential properties of the MW: 
(1) baryonic matter distributed in disk and bulge; 
(2) the MW rotation curve;
(3) gas disk distribution as measured by \citet{Kalberla09};
(4) a hot gaseous halo that matches the core model of \citet{Fang2013}. 
The hot gaseous halo is assumed to be in hydrostatic equilibrium.
Since cooling has been implemented in our simulation code \citep{Wang2012}, 
we set up a hot gaseous halo using the relation between cooling density and temparature for mildly enriched gas, following \cite{Maller2004}.
The initial density profile of the halo hot gas follows the model used by \citet{2002MNRAS.333..481B}: 
$\rho_{\rm halo} \propto (r+a_{\rm halo})^{-4}$,where $r$ is the spherical radius, 
the scaling radius $a_{\rm halo}=400$~kpc in order to match the core hot halo model proposed by \citep{Fang2013}.
The initial properties of the MW model are listed in Table~\ref{tbmw}. 
Though the initial conditions of this MW model, such as the gaseous disk, deviate from the observed properties of the MW, 
this model provides a rather stable galaxy model resembling the MW after the 0.5-Gyr initial relaxation and during the subsequent 2 Gyr.
The mean properties of this MW model, i.e., 1.5-Gyr of evolution in isolation are summarize in Table~\ref{tbmw} (see Fig.~\ref{figmwrc}).
In this paper, we present the simulations with this MW model.
Note that we initially set a hot halo with a mass of $5.3\times10^{10}\Msun$ within the virial radius, i.e., 260 kpc. 
After 1.5-Gyr in isolation, it decreases to $4.8\times10^{10}\Msun$, i.e., a 9\% loss. 
The density distribution of hot gas, as shown in Fig.~\ref{figmwrc}, is comparable to the core model by \citep{Fang2013}.

\begin{table*}
{\centering
\caption{Parameters of the MW model. Note that in order to construct a galaxy model which represent the basic properties of the MW, we artificially input a bugleless disk with large gas extension.}
{\small
\begin{tabular}{rccccccccrr} \hline \hline
Parameters   
& $m_{\rm total}$ 
& $m^{\rm dark}_{\rm halo}$ 
& $m^{\rm hot\,\,gas}_{\rm halo}$ 
& $m_{\rm star + gas\,disk}$
& B/T$^{\rm a}$ 
& $r^{\rm  stellar}_{\rm disk}$ & $r^{\rm  gaseous}_{\rm
  disk}$ &  $f_{\rm gas}$ \\
&  ($10^{10}\Msun$)  & ($10^{10}\Msun$) 
& ($10^{10}\Msun$) 
& ($10^{10}\Msun$) 
&  & (kpc)
& (kpc) & \\ \hline
Initial values($<260{\rm kpc}$)& 75.9 & 64.4 & 5.30  & 5.60  & 0.00  & 2.00 & 8.00 & 0.20 \\
Mean values($<260{\rm kpc}$)& 74.7 & 64.9 & 4.80  & 5.59  & 0.21  & 2.86 & 3.86$^{b}$ & 0.17 \\ \hline
\end{tabular} \\
\label{tbmw}
$^{\rm a}$ B/T: bulge-to-total mass ratio, defined as $m_{\rm bulge}/m_{\rm star}$.
$^{\rm b}$ The scale-length is obtained by fitting the gas density profile from 15 to 30 kpc, in order 
to compare with \citet[][see their fig.\ 5]{Kalberla09}.
}
}
\end{table*}
{\tiny
\begin{table*}
{\centering
\caption{Inital properties of simulated TDGs.}
\label{tbtdg}
\begin{tabular}{l|ccccccccccc} \hline \hline
Name &$m_{\rm gas}$  & $r_{\rm gas}^{\rm half}$$^{\ a}$ & $N_{\rm  gas}$$^{ b}$ 
&$m_{\rm star}$ & $r_{\rm star}^{\rm half}$$^{\ a}$ & $N_{\rm  star}$$^{ b}$ 
& $\sigma^{\rm LoS}_{\rm star}$ & $f_{\rm gas}$ & $M_V$$^{ c}$ &
$(M/L)_{\rm baryonic}$ & $(M/L) _{\rm dyn}$ \\
&($10^6\Msun$) & (kpc) &  & ($10^6\Msun$)  & (kpc) & & ($\kms$) &  & \\ \hline
\multicolumn{12}{l}{Original TDGs taken from the simulated tidal tail} \\ \hline \hline
TDG2 & 483 & 1.55 & $\approx$18k & 101 & 0.78 & $\approx$3700 & 12.1 & 0.82 & -15.1 & 5.8 & 1.4 \\
TDG3 & 162 & 1.14 & $\approx$6000 & 4.9 & 0.50 &  180 & 5.5 & 0.97 & -11.6 & 34 & 7.4 \\
TDG9 & 231 & 1.26 & $\approx$8400  & 18.0 & 0.55  & 655 & 7.4  & 0.93 & -13.0 & 23 & 5.9 \\ 
\hline 
\multicolumn{12}{l}{Resampled TDGs (see text)} \\ \hline \hline 
TDG2 & 470 & 1.56 & $\approx$17k & 133 & 0.80 & $\approx$553k & 12.5 & 0.78 &-15.5& 4.5 & 1.4 \\
TDG3 & 125 & 0.95 & $\approx$4700 & 10.3 & 0.35 & $\approx$419k & 6.7 & 0.92 &-12.5& 13 & 3.1 \\
TDG9 & 230 & 1.24 & $\approx$8500  & 24 & 0.58 & $\approx$429k & 7.8  & 0.90 & -13.5 & 10.5 & 2.3 \\ 
\hline
\end{tabular} \\
$^{a}$ half-mass radius of star and gas, respectively.
$^{b}$ Number of particles. 
$^{c}$ The Absolute V-band magnitude is calculated assuming
a stellar $(M/L)_0=1$ and $M_V^{\rm Sun}=4.8$. 
}
\end{table*}
}

\subsection{Initial conditions of  gas-rich TDGs}
\label{secIC}
\begin{figure}
\includegraphics[width=8cm]{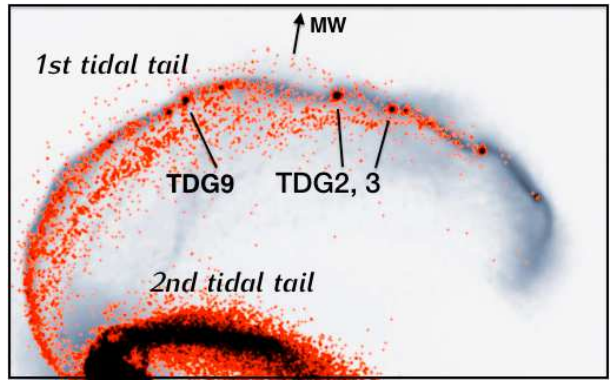}
\caption{Tidal tails in the M31 merger model. 
Stellar particles are shown in red, and gas in grey. 
TDG2, TDG3, and TDG9 are selected from the 1st tidal tail. 
The physical size of the view box is $2\times1.25$ Mpc$^2$.}
\label{figtt}
\end{figure}

Initial conditions of the TDGs are captured from the simulations based on the M31
major merger model from \citet{Hammer2010}. In the simulations, 
we always find  a tidal tail (see Fig.~\ref{figtt}) that is expelled from the secondary galaxy 
along a hyperplane that is close the orbital plane of the merger. 
The initial conditions of the primary and secondary galaxies are set with high gas fractions, 60 and 80\%, respectively 
\citep{Hammer2010} in agreement with expectations from high redshift galaxies \citep{Rodrigues2012}.
It results in very gas-rich TDGs forming in the tidal tails, including the most massive one, TDG2, which shows some 
similarities in location and velocity with the LMC, though its baryonic mass is smaller \citep{Fouquet2012}.
The simulation used here is an improved model over the one of  
\citet{Hammer2010}, which has also been described in \citet{Fouquet2012}.
We selected three TDGs with different masses.
Table~\ref{tbtdg} lists their properties (marked as ``Original TDGs'').  
To investigate the individual TDG initial conditions we extract them with a cube
of 40 kpc$^3$ from the snapshot at 7.5~Gyr after the first passage of the M31 model.
We have carefully checked for dark matter content within the simulated TDGs and found 
 that, within a sphere of 4-kpc radius, the mass of dark matter is on average $\approx4\times10^3\Msun$, which is
negligible compare to the baryonic mass of TDGs.

\begin{figure}
\includegraphics[width=9cm]{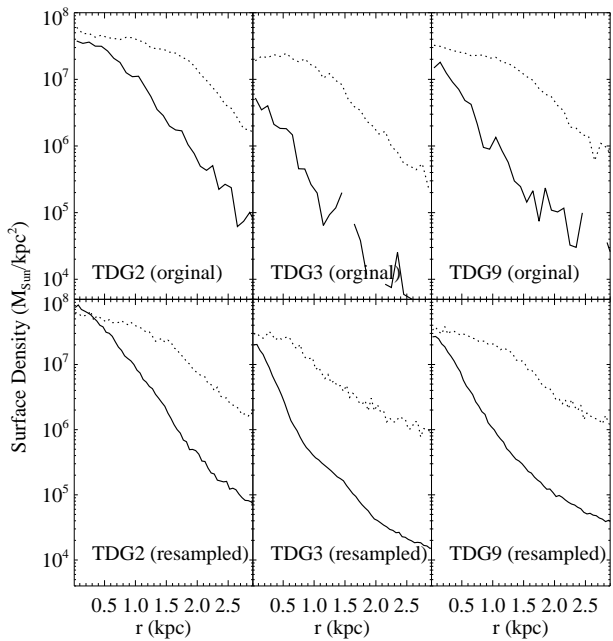}
\caption{Density profiles of simulated TDGs with stellar
components shown as solid lines, and gas as dotted lines.
}
\label{figdenpro}
\end{figure}

The simulation of the M31 model was run with
8 million particles (including 4 million baryon particles), 
nevertheless the resolved TDGs contain a small number of stellar particles, 
as can be seen in Table~\ref{tbtdg}. 
These numbers are  too small to properly determine the TDG properties 
after the passage of their pericenter \citep[see, e.g.,][]{1997NewA....2..139K}.
In order to increase the number of particles, we re-sample
the TDG data in 3D by conserving the density and kinematics. 
The resampling of the TDG ICs is done using several iterations. 
First, we increased the number of particles by a factor of 10, and let the system relax in isolation during 1 Gyr. 
Then, we took the last snapshot and repeat the resampling and relaxation processes several times.
Typically, with 2 to 3 iterations, the total number
of stellar particles can be increased by a factor of several thousands.
We find that the total stellar mass of the TDGs increase by 31\% for TDG2, 33\% for TDG9, 
 when compared to the original TDGs. This is mainly due to the star-formation at the center of the TDGs during the resampling process. 
The resampled TDG3 almost doubled its stellar mass because during the resampling we assumed a low feedback.
Because the star fraction is still small in simulated TDGs, we do not consider this as affecting the overall properties of the simulated TDGs.
Fig.~\ref{figdenpro} shows the density profiles of the TDGs before and after resampling, showing that our procedure has
essentially smoothed the density profiles. 
The above resampling method has been applied only for
stellar particles. We have then assumed different treatments for gas and stars, and have verified that 
with the current number of gas particles, the ram-pressure process can be resolved reasonably well.
Note that in the M31 simulation, we used a softening length of 0.01 kpc for star and gas particles. In current study of TDG-MW interaction, we adopt a larger softening length of 0.05 kpc in the resampling of TDG ICs and simulations in this paper. The change of softening is due to the following reasons: it shortens the computational time allowing a large number of simulations to be tested; with a softening length of 0.05 kpc, the average core radius (about 0.25 kpc) could be well resolved; we have tested the effect induced by the two different softening values and they give almost identical results using our analysis procedure (see the following text).

We derived both of the total $M/L$ and ($M/L$)$_{\rm dyn}$  
for TDG ICs (Table~\ref{tbtdg}). When we estimated the light from stellar mass, 
we assumed a stellar $(M/L_V)_0=1$ as adopted by \citet{2012AJ....144....4M}. 
Because TDGs are gas-rich, their baryonic mass-to-light ratio should account for the mass in gas, i.e.,
$(M/L_V)_{\rm baryonic}=(M/L_V)_0 (m_{\rm star}+m_{\rm gas}) / m_{\rm  star}=(M/L)_0/(1-f_{\rm gas})$, where gas fraction $f_{\rm gas}= m_{\rm  gas}/ (m_{\rm star}+m_{\rm gas})$.
For TDGs with $f_{\rm gas} > 0.9$, this implies $(M/L_V)_{\rm baryonic}>10$ by construction.
Dynamical \mldyn, i.e., Eq.(\ref{eqml}) were also calculated for comparison 
(Table~\ref{tbtdg}). Quantities needed in Eq.(\ref{eqml}) are estimated using the same method described in Sect.~\ref{secres}.
Both types of $M/L$ show lower values than the initial ones, which is due to  the increase of stellar mass.

\begin{table}
\centering{
\caption{Orbits of the TDG-MW collision.}
\label{tborb} 
{\footnotesize
\begin{tabular}{llcc} \hline \hline
Target & Model name$^{\rm a}$ & $r_{\rm peri}$$^{\rm b}$ (kpc)  & Gas removal \\
\hline
 TDG2 
 & TDG2-rp20    & 25.1    &  Yes \\
 & TDG2-rp40    & 58.5    &  No \\
 & TDG2-rp80    & 81.5    &  No \\
 & TDG2-rp100   & 100     &  No \\
TDG3 
 & TDG3-rp20    & 26.5  &  Yes \\
 & TDG3-rp28    & 48.5  &  Yes \\
 & TDG3-rp40    & 45.5  &  Yes \\
 & TDG3-rp80    & 89.5  &  Yes \\
 & TDG3-rp100   & 107   &  Yes \\
 & TDG3-rp200   & 190   &  Yes \\
TDG9 
 & TDG9-rp50    & 44.5  &  Yes \\
 & TDG9-rp60    & 50.5  &  Yes \\
 & TDG9-rp70    & 55.5  &  Yes \\
 & TDG9-rp80    & 57.5  &  Yes \\
 & TDG9-rp100   & 74.5  &  Yes \\
\hline \hline
\end{tabular} \\
$^{\rm a}$ Each model name is composed by a TDG name plus a setting pericenter value in unit of kpc.
$^{\rm b}$ The actual pericenters that are measured from simulations.
}
}
\end{table}
                                                                      
                                                           

\subsection{Orbit of the TDGs relative to the MW}
\label{secorb}
In the present scenario the MW satellites,
including the MCs
originated from a tidal tail (hereafter called TT1) 
which is expelled from M31 during the first passage of the progenitors
\citep{Fouquet2012,Hammer2010,2010ApJ...725L..24Y}.
By tracing TDG2 which resembles the LMC \citep{Fouquet2012}, 
we have found that the orbit of the interaction has a high eccentricity, 
which is calculated when the TDG is approaching the MW at a distance of about 400 kpc, 
and for which the MW mass can be approximated to a point mass.
The high eccentricity results from the combination of the large kinetic energy 
provided by the ancient M31 major merger and of the M31 motion towards the MW at $\approx130\,\kms$.
Due to the internal velocity dispersion in TT1, the orbital eccentricities may 
vary around $e\approx 1.8$. For example, the velocity difference between
TDG2 and TDG3 is about 20~$\kms$, resulting in eccentricities ranging from 1.5 to 2.

We simulated TDG-MW collisions with a fixed orbital eccentricity of 1.8 and different pericenters ranging from 20 to 200 kpc (Table.~\ref{tborb}). The initial position and velocity of each orbit are calculated by assuming a problem of two point masses. Initial orbital parameters are defined when TDG and the MW are separated by 400 kpc, i.e., 1.5 times larger than the virial radius of the MW model.  
All the orbits are set in galacto-centric coordinates \citep{Marel2002}, 
for which the Sun is assumed to be located at $(x,y,z)=(-8.5,0,0)$ kpc.
The resampled TDGs enter from $l\approx90$ degrees and from the south of the Galactic
sphere. 
The orbital plane is set to be 15 degrees inclined to the MW disk 
and crosses the mass center of the MW.
Therefore, the orbits of the TDGs are by construction similar to the trajectory of the LMC 
\citep{2010ApJ...725L..24Y,2007ApJ...668..949B,2006ApJ...638..772K}. 

\begin{figure}
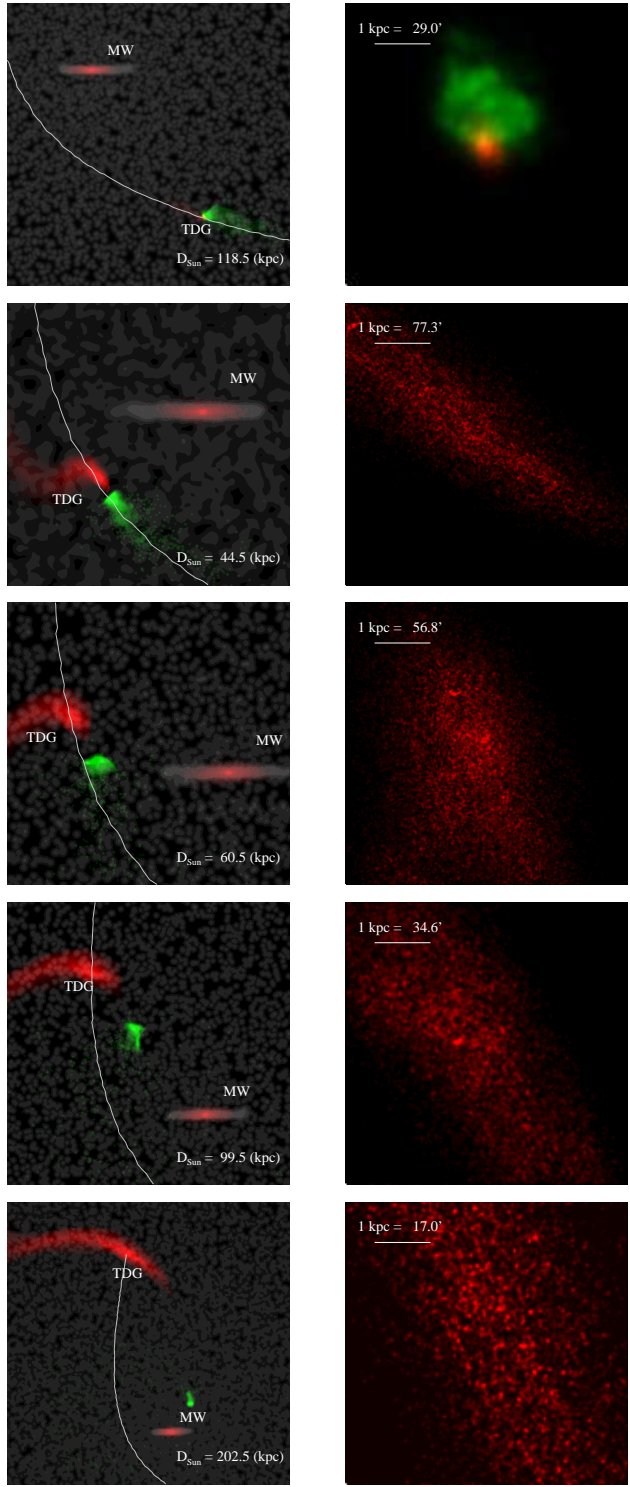

\begin{center}
\includegraphics[width=8.5cm]{\pmora{108}}\\
\includegraphics[width=8.5cm]{\pmora{152}}\\
\includegraphics[width=8.5cm]{\pmora{167}}\\
\includegraphics[width=8.5cm]{\pmora{189}}\\
\includegraphics[width=8.5cm]{\pmora{250}}
\end{center}
\caption{
  Morphological evolution of model TDG9-rp50, 
  illustrating the ram-pressure stripping of the gas and the  tidal disruption of the stars.
  Each snapshot is shown with two panels: a wide view on the left with heliocentric distance of TDG9 indicated at the bottom-right,
  and a close-up look from the Sun on the right.
  The white lines indicate the trajectory of the TDG. 
  Stars are encoded in red, MW gas in grey, TDG gas in green. The contrast between different
  components is arbitrary. 
  To capture the TDG evolution up to a very large distance, 
  we used a detection limit of $30\,{\rm mag}\,{\rm arcsec}^{-2}$.
}
\label{figmor}
\end{figure}

\section{Results}
\label{secres}
We captured snapshots every 10 Myr for each simulation.
In order to compare with observations, we defined a way to 
estimate the observable quantities of TDGs from the simulations. 
First, we projected TDG stellar particles on celestial coordinates, 
i.e., assuming that the observer is at the Sun location.
The object detection is performed in this frame.
Assuming a stellar $(M/L_V)_0=1$,
the stellar mass of each particle can be converted into V-band luminosity 
by taking the absolute magnitude of the Sun $M_V=4.8$. 
Using a sampling size of $\Delta r=0.1$\,kpc, 
which is twice the softening length used in the simulations, 
we derived light (i.e., mass) profiles and velocity dispersion profiles. 
Central velocity dispersions ($\sigma_0$) are estimated using a 0.2-kpc aperture around the center of the TDGs.
Light profiles are fitted using a S\'{e}rsic function, from which
central surface brightness ($\mu_0$) and core radius (i.e., half-brightness radius $r_{\rm c}$) are derived.
\citet{1995MNRAS.277.1354I} described 
observational detection limits due to the foreground MW star counts. They vary from one object to another, 
with a mean value of 27~mag~${\rm arcsec}^{-2}$. 
We adopted this detection limit to define an observable region within which 
the absolute magnitudes of TDGs can be derived.
The gas fraction, $f_{\rm gas}(< 4\,\rm{kpc})$, is defined to be the mass ratio of gas to  total 
within a sphere of 4-kpc radius at the TDG location. 
In the following sub-sections, we describe in detail the evolution of TDG properties, such as mass, morphology, kinematics, and \mldyn\
in some representative models. The \mldyn\ is calculated according to Eq.~\ref{eqml} 
with the necessary quantities estimated above.

\subsection{Morphological Evolution}
Fig.~\ref{figmor} shows the evolution of model TGD9-rp50. 
From the wide view panels, one may notice that the ram pressure works efficiently in removing gas from the TDG.
When the TDG approaches its pericenter (see the second row), 
gas and stars have been fully decoupled. In the following panels, the gas significantly drifts from the orbit of stars and goes towards the MW center, indicating a strong orbital energy dissipation. 
After passing the pericenter, 
the stripped gas falls back towards the MW disk, 
while stars run farther since they do not feel any more the ``internal friction'' 
(due the gravity between stars and gas when the gas is stripped) caused by ram pressure. 
The close-up views of Fig.~\ref{figmor} show how morphologies would be observed
at different epochs.  The elongated shape of the
galaxy after pericenter passage results form the impact of the tidal stirring.

\begin{figure}
\includegraphics[width=8cm]{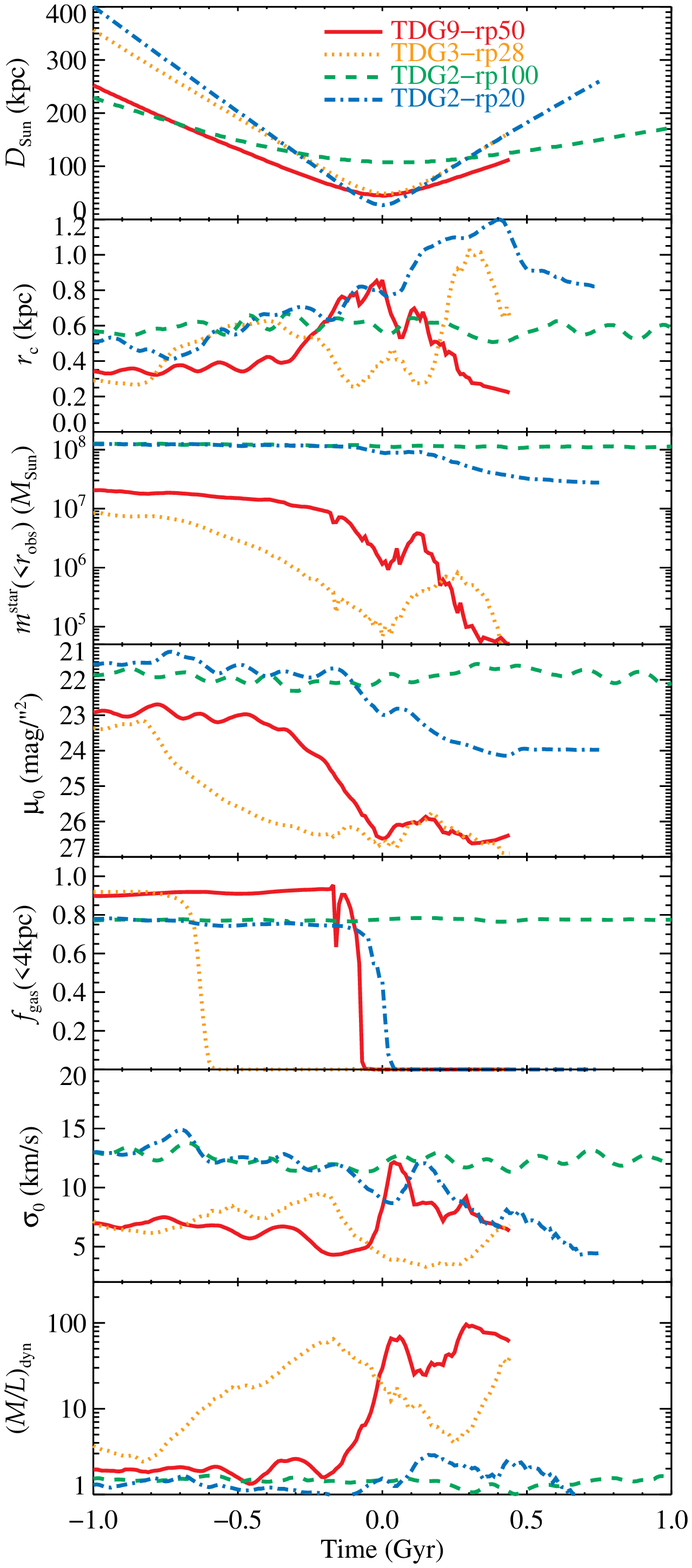}
\caption{
Evolution of TDG properties for four
models that are indicated in the top panel. 
For all the simulations, the reference time is set to be $T=0$ when TDGs arrive to their pericenters.
From top to bottom, panels are
heliocentric distance ($D_{\rm Sun}$), 
core radius ($r_{\rm c}$), 
stellar mass enclosed within the observable radius $r_{\rm obs}$, 
central surface brightness ($\mu_0$), 
gas fraction ($f_{\rm gas}$) within a sphere of 4-kpc radius,
central velocity dispersion ($\sigma_0$),
and $(M/L)_{\rm dyn}$. Some of the evolution tracks are stopped before the end of the simulations because the objects are fainter than our detection limit of  $27\,{\rm mag}\,{\rm arcsec}^{-2}$.
}
\label{figml}
\end{figure}

Fig.~\ref{figml} demonstrates the evolution of the morphological properties 
obtained from four simulations: TDG9-rp50, TDG3-rp28, TDG2-rp100 and TDG2-rp20. 
For all TDG2 simulations, we find that its properties are marginally affected by the ram-pressure and
tidal forces exerted by the MW, except when it passes at a small pericenter of $25$ kpc (Fig.~\ref{figml}). This is because TDG2 is too massive to be affected. On the contrary, TDG3 and TDG9 structures are significantly altered.
Compared to their initial values,  TDG3 and TDG9 core radii
may increase by  factors ranging from 2 to 3.

The stellar masses enclosed within $r_{\rm obs}$ of TDG3 and TDG9 dramatically decrease
by factors larger than 100, an evolution that is linked to the adopted detection limits, as actually in real observations.
A strong decrease of the central surface brightness, by up to 4 magnitudes, 
suggests that a strong dilution of the central star density, implying that
the TDG stellar content is expanding at all scales. 
Looking at the evolution of the gas fraction, we further notice that
the expansion process is triggered by the gas removal.

\subsection{Flat velocity dispersion profiles}
Fig.~\ref{figml} shows the evolutions of $\sigma_0$, 
i.e., the central velocity dispersion. As we mentioned above, 
the TDG2 properties are marginally affected.
TDG3 and TDG9 have a moderate evolution of $\sigma_0$. The largest variations (by a factor 2)
are provided by the passages near their pericenters.

\begin{figure}
\includegraphics[width=8cm]{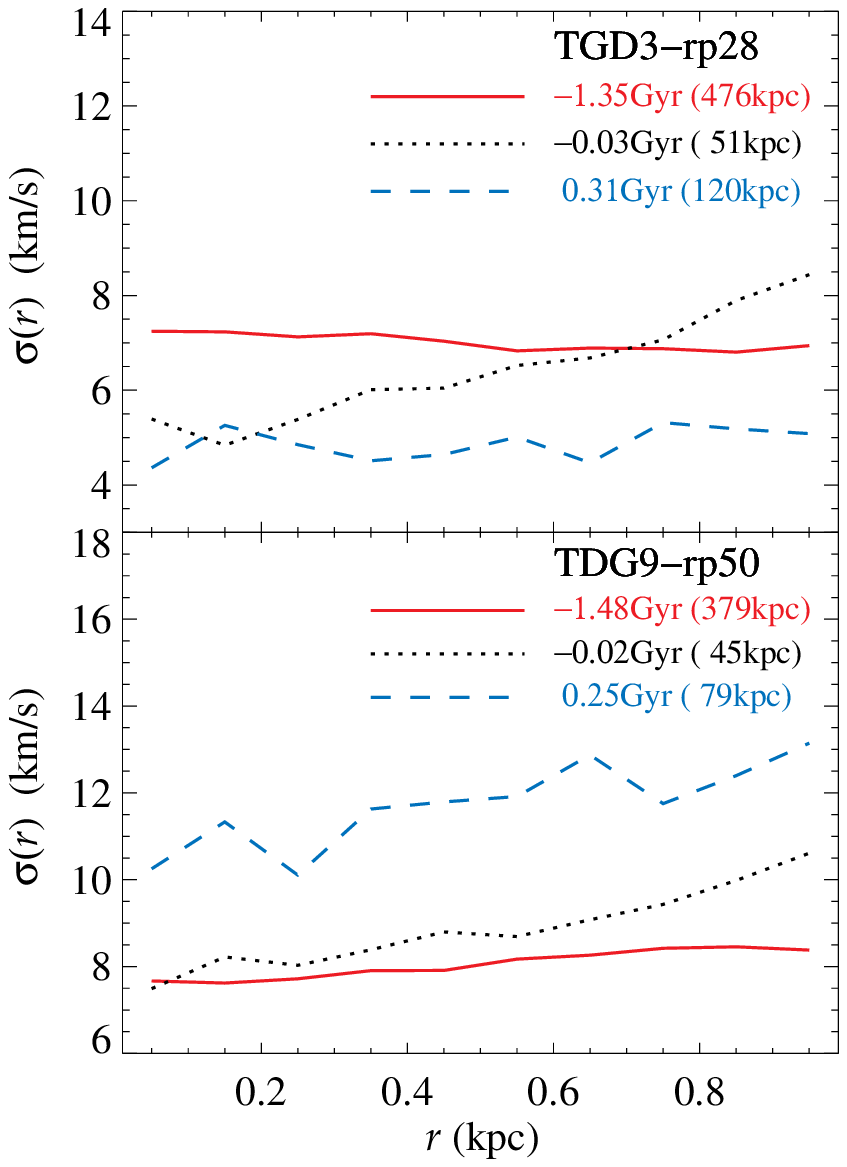}
\caption{Evolution of the velocity dispersion profiles for two simulations. 
For each simulation, we selected three representative profiles at different times, 
i.e., before, during and after pericenteric passage.
The heliocentric distances are also indicated for different epochs.
}
\label{figsigprof}
\end{figure}

The evolution of the dispersion profile for two models is shown in Fig.~\ref{figsigprof}. 
Both TDG3 and TDG9 at initial conditions (seeded lines) show an almost flat profile.
 Gas stripping  and tidal stirring at pericenter provide a slightly increasing
velocity dispersion profile with radius.
This result matches the observed flat or increasing dispersion profiles of the MW dSphs 
\citep[e.g.,][]{2009ApJ...704.1274W}.
Therefore, our simulations indicate that the flatness of the dispersion profile of the MW dSphs 
could be associated to interactions between their progenitors, here gas-rich
TDGs, and the MW.

\subsection{\mldyn\  Evolution} 
The bottom panel of Fig.~\ref{figml} shows 
the evolution of $(M/L)_{\rm dyn}$ according to Eq.(\ref{eqml}).
Models TDG3 and TDG9 show that $(M/L)_{\rm dyn}$ may increase dramatically 
from a few to 100 or more, within less than 1~Gyr of interaction with the Milky Way.
Because the observed $r_{\rm c}$ and $\sigma_0$
may change by factor of about 2-3, respectively, these parameters
are not driving the $(M/L)_{\rm dyn}$ evolution.
The latter is mostly caused 
by the considerable decrease of central surface brightness ($\mu_0$) that is 
associated to the rapid expansion of the initial stellar distribution. 
It implies that strong perturbations caused by the MW almost destroy the TDG remnants,
and that most of the apparent high $(M/L)_{\rm dyn}$ is due to the decrease of $\mu_0$.

\begin{table*}
\caption{Observed internal properties of the MW satellites.}
\label{tbdwobs}
{\footnotesize
\begin{tabular}{lr@{$\pm$}lr@{$\pm$}lr@{$\pm$}lr@{$\pm$}lr@{$\pm$}lr@{}llll} \hline \hline
Name & 
\multicolumn{2}{c}{$M_V$} &  
\multicolumn{2}{c}{$r_{\rm c}$} &
\multicolumn{2}{c}{$r_{\rm h}$}&  
\multicolumn{2}{c}{$\sigma_0$}& 
\multicolumn{2}{c}{$\mu_0$}&  
\multicolumn{2}{c}{{\mldyn}}  &  $M_{\rm H I}$ & $M_{\rm stellar}$ & $f_{\rm gas}$$^{\rm a}$\\ 
&
\multicolumn{2}{c}{} &  
\multicolumn{2}{c}{(pc)} &
\multicolumn{2}{c}{(pc)}&  
\multicolumn{2}{c}{($\kms$)}& 
\multicolumn{2}{c}{${\rm mag}/\prime\prime^2$ }&  
\multicolumn{2}{c}{}  & ($10^6\Msun$) & ($10^6\Msun$) &  (\%) \\  \hline
Sculptor      &-11.1 &0.3  &110 &30 &260& 39  &6.6  &0.7  &23.7 &0.4 & 11 &$^{+27}_{-5}$   &  0$^{\rm a}$     & 2.3 &  0 \\
Fornax        &-13.2 &0.3  &460 &27 &668& 34  &10.5 &1.5  &23.4 &0.3 & 5  &$^{+9}_{-3}$    & $>$0.146$^{[2,3]}$   & 20  &  0.73       \\
Carina        & -9.3 &0.25 &210 &29 &241& 23  &6.8  &1.6  &25.5 &0.4 & 33 &$^{+84}_{-12}$  & $<$0.0002$^{[2]}$    & 0.38&  0.053      \\
Leo I         &-11.9 &0.3  &215 &20 &246& 19  &8.8  &0.9  &22.4 &0.3 & 3  &$^{+5}_{-2}$    & $<$0.0015$^{[2]}$  & 5.5 &  0.027       \\
Sextans       & -9.5 &0.3  &335 &24 &682& 117 &6.6  &0.7  &26.2 &0.5 & 37 &$^{+77}_{-17}$  & $<$0.00018$^{[2]}$  & 0.44&  0.041   \\
Leo II        & -9.6 &0.2  &160 &33 &151& 17  &6.7  &1.1  &24.0 &0.3 & 11 &$^{+24}_{-5}$   & $<$0.01$^{[2]}$     & 0.74&  1.3       \\
UMi    & -8.9 &0.4  &200 &15 &280& 15  &9.3  &1.8  &25.5 &0.5 & 65 &$^{+158}_{-25}$ & $<$0.04$^{[2]}$     & 0.29& 12.    \\
Draco         & -8.8 &0.3  &180 &14 &196& 12  &9.5  &1.6  &25.3 &0.5 & 62 &$^{+146}_{-25}$ & $<$0.00016$^{[2]}$  & 0.29&  0.055   \\
Sagittarius   &-13.4 &0.5  &550 &46 &\multicolumn{2}{c}{-}  &11.4 &0.7  &25.4 &0.3 & 32 &$^{+52}_{-20}$ & $<$0.00014$^{[2]}$  & 21  &  0.001 \\
SMC           &-16.2 & 0.2$^{[4]}$ &\multicolumn{2}{c}{-} &\multicolumn{2}{c}{-} & \multicolumn{2}{c}{-} & \multicolumn{2}{c}{-} &  \multicolumn{2}{c}{-} &   402$^{[2]}$          & 460$^{[4]}$ & 46.6    \\
LMC           & -18.1& 0.1$^{[4]}$ &\multicolumn{2}{c}{-} & \multicolumn{2}{c}{-} &\multicolumn{2}{c}{-}& \multicolumn{2}{c}{-}  & \multicolumn{2}{c}{-} &   500$^{[2]}$          & 1500$^{[4]}$& 25.  \\
\hline \hline                                                                                            
\end{tabular} 
}                       \\                                                                    
Note that the data are collected from \citet{1998ARA&A..36..435M} except mentioned otherwise: 
$^{[1]}$ \citet{2003AJ....126.1295B},
$^{[2]}$ \citet{2009ApJ...696..385G},
$^{[3]}$ \citet{2006AJ....131.2913B},
$^{[4]}$ \citet{2012AJ....144....4M}.
$^{\rm a}$ Gas fractions are corrected for the mass fraction of hydrogen which is 0.76. Note that for Sculptor we set its H$\,${\sc i} mass and gas faction to be zero because it has no significant star-formation over the past 7 Gyr \citep{deBoer2012}.
\end{table*}

\begin{table*}
\caption{Observed dynamical properties of the MW satellites.}
\label{tbdwdyn}
{\footnotesize
\begin{tabular}{lr@{, }lr@{$\pm$}lr@{$\pm$}lr@{$\pm$}lr@{$\pm$}lr@{$\pm$}l} 
\hline \hline
Name & \multicolumn{2}{c}{($l',b'$)} &  \multicolumn{2}{c}{ $D_{\odot}$ (kpc) } & \multicolumn{2}{c}{ $V_{\rm r}$  (km/s) } &\multicolumn{2}{c}{ $\mu_{\rm W}$}($\masyr$)& \multicolumn{2}{c}{$\mu_{\rm N}$ } ($\masyr$)\\ \hline
Sculptor    &287.5 & $ -83.2$  &  79 &  4&  $ 108.0$&3  &   $-0.090$ &  0.013&  $ 0.020$ &  0.013$^{[2]}$  \\
Fornax      &237.1 & $ -65.7$  & 138 &  8&  $  53.0$&3  &  $-0.476$ &  0.046&  $-0.360$ &  0.041$^{[3]}$  \\
Carina      &260.1 & $ -22.2$  & 101 &  5&  $ 224.0$&3  &  $-0.220$ &  0.090&  $ 0.150$ &  0.090$^{[4]}$  \\
Leo I       &225.9 & $  49.1$  & 250 & 30&  $ 286.0$&2  &  $ 0.114$ &  0.030&  $-0.126$ &  0.029$^{[5]}$  \\
Sextans     &243.5 & $  42.3$  &  86 &  4&  $ 227.0$&3  &  $-0.260$ &  0.410&  $ 0.100$ &  0.440$^{[6]}$  \\
Leo II      &220.2 & $  67.2$  & 205 & 12&  $  76.0$&2  &  $-0.096$ &  0.105&  $-0.033$ &  0.151$^{[7]}$  \\
UMi  &105.0 & $  44.8$  &  66 &  3&  $-248.0$&2  &  $ 0.500$ &  0.170&  $ 0.220$ &  0.160$^{[8]}$  \\
Draco       & 86.3 & $  34.7$  &  82 &  6&  $-293.0$&2  &  $-0.101$ &  0.069&  $-0.030$ &  0.120$^{[9]}$  \\
Sgr         &  5.5 & $ -14.2$  & 24 &   2&  $ 140.0$&5  &  $ 2.560$ &  0.080&  $-0.880$ &  0.080$^{[10]}$  \\
SMC         &302.8 & $ -44.3$  & 64 & 4$^{[1]}$  &  $ 145.6$&0.6$^{[1]}$  &  $-0.980$ &  0.290&  $-1.010$ &  0.300$^{[11]}$  \\
LMC         &280.5 & $ -32.5$  & 51 & 2$^{[1]}$  &  $ 262.2$&3.4$^{[1]}$  &  $-1.890$ &  0.270&  $ 0.390$ &  0.270$^{[11]}$  \\
\hline \hline
\end{tabular} \\
Note that the data of ($l',b'$), $D_{\odot}$, and $V_{\rm r}$ are collected from \citet{1998ARA&A..36..435M} unless mentioned: $^{[1]}$ \citet{2012AJ....144....4M}. 
References for proper motions: $^{[2]}$\citet{Piatek2006}, $^{[3]}$\citet{Piatek2007}, $^{[4]}$\citet{Piatek2003}, $^{[5]}$\citet{Sohn2013}, $^{[6]}$\citet{Walker2008}, $^{[7]}$\citet{Lepine2011}, $^{[8]}$\citet{Piatek2005}, $^{[9]}$\citet{Piatek2008}, $^{[10]}$\citet{Ibata1997}, and $^{[11]}$\citet{Vieira2010}.
}
\end{table*}
\begin{table*}
\caption{Derived orbital parameters of the MW satellites.} 
\label{tbdworb}
{\footnotesize
\begin{spacing}{1.2}
\begin{tabular}{lr@{}lr@{}lr@{}lr@{}lr@{}l} \\ \hline \hline
Name & \multicolumn{2}{c}{$r_{\rm p}$  (kpc)} &  \multicolumn{2}{c}{$r_{\rm a}$ (kpc)} &   \multicolumn{2}{c}{$P$  (Gyr)} &  \multicolumn{2}{c}{$e$ } &  \multicolumn{2}{c}{$T_{\rm rp}$ (Gyr)}\\ \hline
Sculptor  & $ 68.2$&$_{ -1.2}^{+  1.2}$ & $176.7$&$_{-10.0}^{+ 11.0}$ & $ 3.82$&$_{-0.25}^{+ 0.25}$ & $ 0.44$&$_{-0.02}^{+ 0.02}$ & $-0.25$&$_{-0.01}^{+ 0.02}$ \\
Fornax    & $135.7$&$_{ -5.4}^{+  2.6}$ & $223.3$&$_{-68.4}^{+132}$ & $ 6.24$&$_{-2.00}^{+ 2.80}$ & $ 0.29$&$_{-0.10}^{+ 0.18}$ & $ 0.27$&$_{-0.14}^{+ 0.26}$ \\
Carina    & $ 25.4$&$_{-19.3}^{+ 20.7}$ & $106.8$&$_{ -1.7}^{+  1.3}$ & $ 1.78$&$_{-0.31}^{+ 0.29}$ & $ 0.57$&$_{-0.23}^{+ 0.21}$ & $-0.79$&$_{-0.12}^{+ 0.08}$ \\
Leo I     & $ 80.8$&$_{-32.9}^{+ 32.1}$ & $ 956.8$&$_{-204}^{+  76}$ & $25.68$&$_{-2.84}^{+ 1.96}$ & $ 1.51$&$_{-0.23}^{+ 0.23}$ & $-1.04$&$_{-0.01}^{+ 0.02}$ \\
Sextans   & $ 86.2$&$_{ -6.3}^{+  1.7}$ & $131.7$&$_{-36.5}^{+194}$ & $ 3.04$&$_{-1.40}^{+ 4.20}$ & $ 1.00$&$_{-0.65}^{+ 0.15}$ & $-0.09$&$_{-0.41}^{+ 0.07}$ \\
Leo II    & $207.2$&$_{ -1.2}^{+  0.8}$ & $251.3$&$_{-31.9}^{+248}$ & $ 7.66$&$_{-3.88}^{+ 7.12}$ & $ 1.02$&$_{-0.85}^{+ 0.15}$ & $-0.06$&$_{-0.07}^{+ 0.03}$ \\
UMi & $ 41.5$&$_{-16.7}^{+ 19.3}$ & $ 82.8$&$_{ -5.0}^{+  9.0}$ & $ 1.52$&$_{-0.42}^{+ 0.58}$ & $ 0.38$&$_{-0.06}^{+ 0.14}$ & $ 0.38$&$_{-0.03}^{+ 0.03}$ \\
Draco     & $ 53.8$&$_{ -9.4}^{+  8.6}$ & $136.1$&$_{-20.4}^{+ 27.6}$ & $ 2.72$&$_{-0.54}^{+ 0.66}$ & $ 0.48$&$_{-0.03}^{+ 0.03}$ & $ 0.39$&$_{-0.05}^{+ 0.02}$ \\
Sgr       & $ 12.8$&$_{ -0.2}^{+  0.2}$ & $ 41.1$&$_{ -2.1}^{+  2.4}$ & $ 0.64$&$_{-0.03}^{+ 0.03}$ & $ 0.52$&$_{-0.02}^{+ 0.00}$ & $-0.04$&$_{-0.00}^{+ 0.00}$ \\
SMC       & $ 38.9$&$_{-22.3}^{+ 17.7}$ & $ 63.5$&$_{ -0.3}^{+  0.7}$ & $ 1.26$&$_{-0.32}^{+ 0.28}$ & $ 0.20$&$_{-0.14}^{+ 0.22}$ & $-0.49$&$_{-0.07}^{+ 0.09}$ \\
LMC       & $ 48.5$&$_{ -1.2}^{+  0.8}$ & $ 224.2$&$_{-152}^{+ 209}$ & $ 4.38$&$_{-2.94}^{+ 4.86}$ & $ 1.04$&$_{-0.39}^{+ 0.16}$ & $-0.04$&$_{-0.03}^{+ 0.01}$ \\
\hline \hline                         
\end{tabular} \\
\end{spacing}
Columns are object name, pericenter ($r_{\rm p}$), aprocenter ($r_{\rm a}$), orbital peroid ($P$), 
orbital eccentricity ($e$), and the nearest time to the pericenter ($T_{\rm rp}$, 
negative values indicate the objects have passed their pericenter). Note that $r_{\rm a}$ is calculated only for the solutions with $e<1$.
}
\end{table*}

\begin{figure*}
\includegraphics[width=14cm]{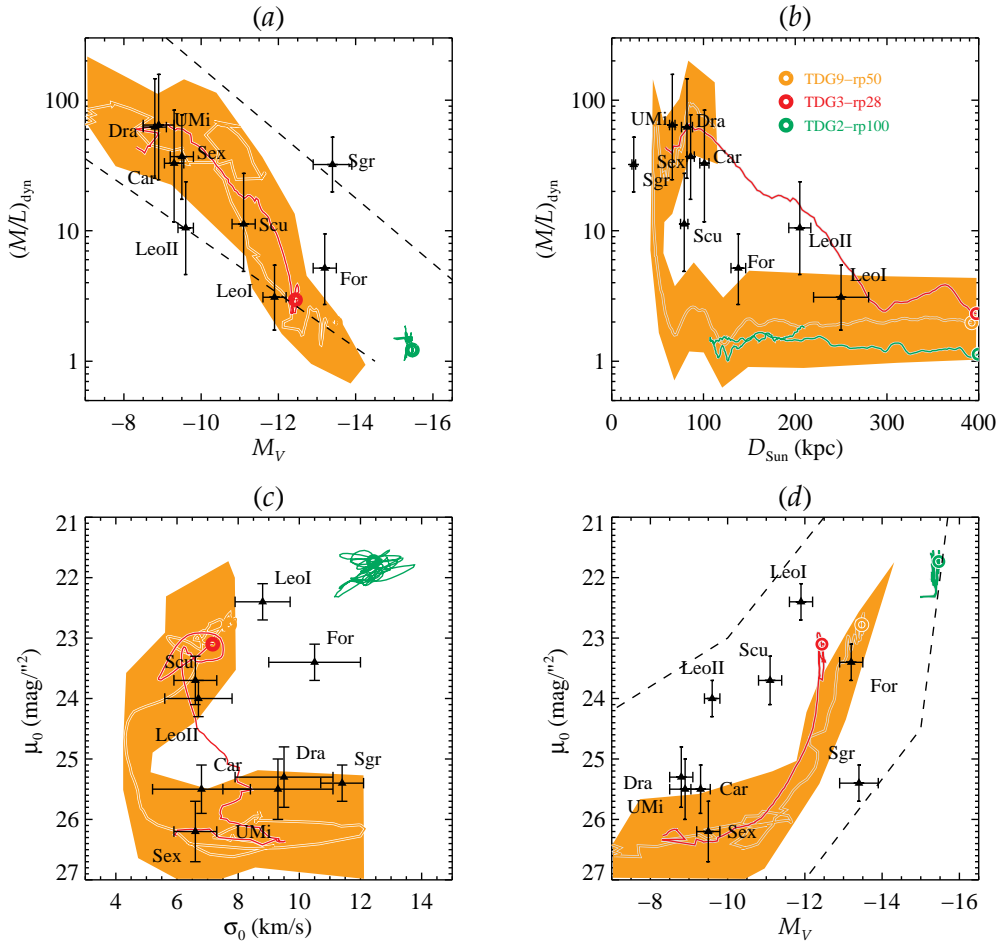} 
\caption{Comparison of simulation results with the observational properties of MW dSphs.  
  In all panels, the properties of the MW dSphs 
  are indicated with black filled triangles with error bars attached.
  Three models (see annotations in panel $b$) are shown with different colors.
  Each evolution track starts with a circle from which the TDG is approching the MW for the first time.
  The dashed lines in panel ($a$) and ($d$) indicate the observational trend of the Local Group dwarf galaxies \citep{2012AJ....144....4M}. 
  The orange area represents the impact on the model TDG9-rp50 of an uncertainty of $\pm 0.3$ dex in the stellar mass-to-light ratio. The same broadness of uncertainty regions caused by stellar mass-to-light ratio may apply to the other models.
}
\label{figsimobs}
\end{figure*}

\section{Could dark-matter free progenitors explain the large \mldyn\  of  dwarf spheroidals?}

We have investigated the interactions between simulated, gas-rich TDGs and the MW.
The simulations suggest a possible mechanism to explain the apparently large \mldyn\ (see Table~\ref{tbdwobs}) values of MW dSphs, during a process that is
 quite similar to former attempts for explaining the dIrr to dSph transformation. 
\cite{2013ApJ...764L..29K} have presented a description of the tidal stirring effect for dSph progenitors
assumed to be dark matter dominated. They convincingly showed that for a shallow enough dark matter
profile, tidal stirring does not only distort the galaxy shape but may transform a rotationally supported dIrr
into a dSph. \cite{2010AdAst2010E..25M} showed that during such a process most of the initial gas mass is lost.

Here, we show that dark-matter free  progenitors such as gas-rich TDGs may provide a similar behavior. When passing through the MW hot gaseous halo, the less massive ones, such as TDG3 and TDG9, may loose their gas 
which corresponds to 90\% of their initial mass. 
Gas removal due to the ram pressure from the MW gaseous halo and tidal stirring due to the disk are
facilitated by the absence of dark matter together with the large orbital eccentricity
implied by our scenario, i.e., a tidal tail coming from M31 at large speed consistent with the
observed value for the LMC.
At the first passage a TDG may become a gas-poor dSph. 
Stars in TDG3 become free as they become decoupled from their dominant mass component, the gas,
and as a consequence, they are expanding steadily.

Eq.~\ref{eqml} is currently used for estimating \mldyn, which includes
three quantities, $r_{\rm c}$, $\sigma_0^2$ and $\mu_0$. 
During the process, core radius and central velocity dispersion do change, 
but only by modest factors (see Fig.~\ref{figml}).  
As shown by the model of TDG9-rp50 in Fig.~\ref{figsimobs} (panels $a$ and $c$), we confirmed the results by \citet{1997NewA....2..139K,2007MNRAS.374.1125M,2012MNRAS.424.1941C} that when a TDG is tidally disrupted
its large non-isotropic velocity dispersion may boost the \mldyn. 
Here, we also find that the large \mldyn\ is mainly caused by the dramatic decrease of $\mu_0$ by 3-4 magnitudes due to
tidal stirring of very fragile objects having lost most of their baryonic content. 
It is very encouraging that our results are matching well the behavior of dSphs in $M_V$-\mldyn and $M_V$-$\mu_0$ planes \citep{1998ARA&A..36..435M,2012AJ....144....4M} as shown by Fig.~\ref{figsimobs} panels $a$ and $d$. This supports the fact that large \mldyn\ is correlated with lower surface brightness (Fig.~\ref{figmuml}).  
Sgr appears as an outlier in Fig.~\ref{figsimobs} panels $a$ and $d$, but well correlated in Fig.~\ref{figmuml}. We find that its large magnitude may be linked to its close distance (see more details in Sect.\ref{secfit}). Our simulations reproduce quite well the already noticed trend for dSphs \citep[][see their fig.\ 11]{2012AJ....144....4M}, i.e., that the most luminous dSphs
are those with the smallest \mldyn. This leads us to investigate
how actual dSph properties could be reproduced using a limited number of simulations.

\section{Attempt to fit individual MW dSph properties}
\label{secfit}

\subsection{Orbital parameters of MW dwarfs}

To compare our limited modeling to actual dSphs requires {\it a priori} knowledge of their orbital properties that can be deduced from their estimates of distances and proper motions.
\citet{2010ApJ...725L..24Y} have presented an analytical model of the MW potential with which we are able to investigate the orbital parameters for each MW satellite.
The total mass of the MW model in \citet{2010ApJ...725L..24Y} is about $1\times10^{12}\Msun$ while in the current paper, the N-body model for the MW has a total mass of $7.5\times10^{11}\Msun$ (Table~\ref{tbmw}). Both models fit well the rotation curve of the MW. The difference in total mass is caused by the different halo profiles, a NFW profile in \citet{2010ApJ...725L..24Y} and a core model in this paper \citep[see][]{2002MNRAS.333..481B}. Both models give a similar distribution of bound/unbound orbits for the MW dSphs, so only one table of solutions with the more massive MW model is presented.
The observed distances, proper motions and radial velocities are listed in Table~\ref{tbdwdyn} and 
derived orbital parameters are listed in Table~\ref{tbdworb}. 
We have performed  Monte Carlo simulations to derive the orbital parameter error
distribution assuming that they are dominated by proper
motion uncertainties. For each dwarf galaxy, we sampled 3000 possible
proper motions using a Gaussian distribution.
Following \citet{Fouquet2012}, we adopted the a proper motion for the LMC from \citet{Vieira2010}
instead of the latest result by \citet{K13}. The latter leads to a solution of $r_{\rm p}=48.5\pm0.1$~kpc, $r_{\rm a}=1176.5_{-387.9}^{+ 792.1}$, $e=0.93\pm0.02$, and $P=35\pm15$~Gyr.
Both orbital solutions using \citet{Vieira2010} or \citet{K13} favor first in-fall scenario of the LMC, 
and lead to the same conclusions in the following discussion.

According to its large orbital period, it is likely that Leo I has experienced only a single passage from the MW about 1 Gyr ago
\citep[see also][]{Mateo2008,Sohn2013,Boylan2013}.
The solutions of Sextans, Leo~II and LMC peak at parabolic orbits, while
Sextans and Leo~II orbit determinations are affected by their large errors in proper motion leading to relatively large uncertainties in orbital eccentricity. 
Sculptor, Fornax, Carina, UMi, Draco, Sgr, SMC are on bound orbits with $e$ ranging from 0.2 to 0.6 and a typical error of 0.13. 
The Sgr orbit is well determined with a 0.64~Gyr period and should have experienced at least one full orbital trajectory since its debris is observed over the whole sky \citep{Majewski2003}.  Pericenters of the MW satellites are distributed in a large range from 12 to 200 kpc, with a median value of 54 kpc.  This large median pericenter gives strong constraints on any model that attempts to explain the MW dSph formation through gas stripping, since this process should work at such a large distance.

\begin{figure}
\includegraphics[width=8cm]{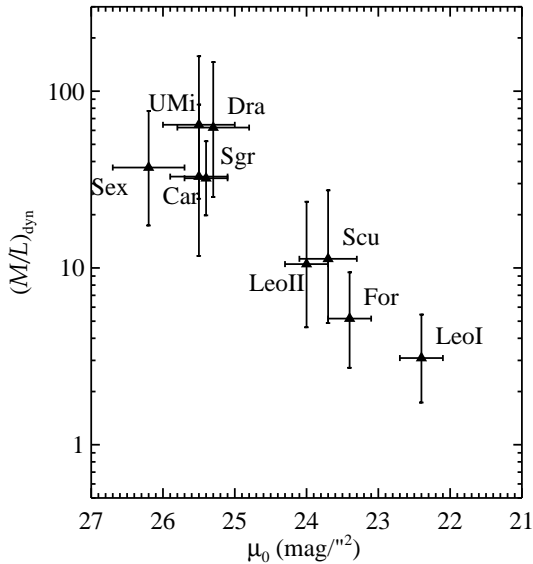} 
\caption{Correlation between centeral suface brighness and dynamical M/L for the MW dSphs. 
}
\label{figmuml}
\end{figure}

\begin{figure}
\includegraphics[width=8cm]{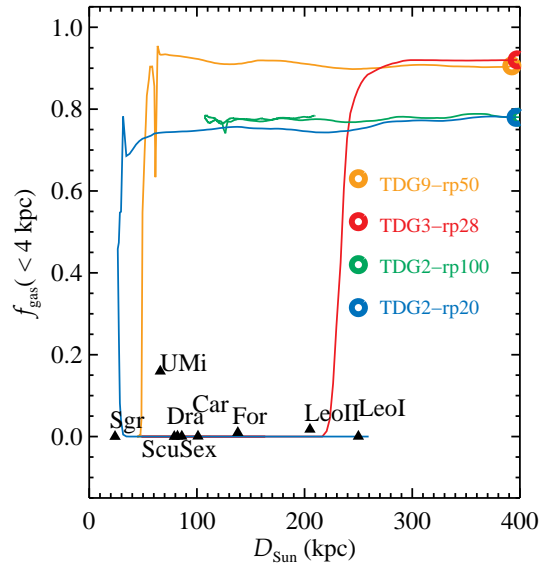} 
\caption{Comparison between the evolution of the gas fraction in the simulations and observations as a function of heliocentric distance.
}
\label{figsimgas}
\end{figure}

\subsection{Fitting of individual MW dwarfs with TDG models}
\label{secfit}
\subsection*{From Gas-rich TDGs to dSphs}
Fig.~\ref{figsimgas} shows gas-fraction evolution of TDGs. Due to the ram pressure from the MW hot halo, TDGs gas could be removed completely in the approaching phase, e.g., for the model TDG3-rp28 the gas is fully stripped 200 kpc before approaching pericenter. Fig.~\ref{figsimgas} also evidences that gas removal by ram-pressure is increasingly efficient for decreasing TDG masses. 
Massive TDG2 may hold their gas except for small pericenter of 25 kpc. 
Table~\ref{tborb} indicates for each simulation whether the TDG gas is completely stripped during a single passage.
Ram-pressure stripping is always efficient in transforming gas-rich TDG3 and TDG9 into gas deficient galaxies during a single passage, even for large pericenters up to 200 kpc.

The gas-striping process strongly affects the apparent properties of the TDG residuals. 
For example, in Fig.~\ref{figml} at $T>0$, TDG3 evolves towards smaller velocity dispersions while TDG9 evolves towards larger velocity dispersions.
This is because they are at different stages of evolution after they completely lost their gas by ram-pressure stripping.
TGD3 has completely lost its gas at 200 kpc when it was approaching the MW, 
corresponding to $T=-0.5$~Gyr in Fig.~\ref{figml}. Then, TDG3 enters an expanding process, leading to a continuous decrease of central surface brightness. During this expanding stage, its velocity dispersion increases (Fig.~\ref{figml}). After TGD3 passes the pericenter ($T=0$), its stellar component starts to concentrate because of gravity, forming a gasless object, which can be evidenced by its mass and central surface brightness evolution.  As this gasless object has much smaller mass ($\approx 10^6\Msun$) compared to its initial total mass ($\approx 10^8\Msun$, see Table.\ref{tbtdg}), we observed a smaller velocity at this stage with respect to its initial value.
TDG9, which is more massive than TDG3, has completely lost its gas later than TDG3, almost at pericenter. During $T=0$~-~0.5 Gyr (Fig.~\ref{figml}), TDG9 is in the stage of expanding, thus increasing its velocity dispersions.

\subsection*{LMC \& SMC}
The LMC-SMC is possibly an interacting pair \citep{2010ApJ...721L..97B} that is dominated by the LMC.
Our scenario that the MW satellite system originated from a M31 tidal tail 
\citep{2010ApJ...725L..24Y,Fouquet2012} is built on the basis of the large motion of the LMC.
Thus, the LMC and SMC are by construction consistent with our models, as their gas-rich content is
fully in agreement with the fact that they are more massive than TDG2, and the expectation that they are on their first passage at pericenter (see Table~\ref{tbdwdyn}). 
As shown in Fig.~\ref{figml} and Fig.~\ref{figsimobs}, a massive object, such as TDG2, may survive and keep its gas close to its pericenter, at 50 kpc from the MW center. 
A discussion of the proposed interpretation of the LMC as a massive TDG can be found in \citet{2010ApJ...725L..24Y}, \citet{Fouquet2012}, and \citet{Hammer2013}.

\subsection*{UMi, Draco, Sextans, and Carina}
Fig.~\ref{figsimobs} shows that these four galaxies share quite similar properties, within quite large error bars. Their properties are well consistent with TDG9-rp50 (orange curve) after it passes pericenter. Here, we consider an uncertainty of $\pm 0.3$ dex in stellar mass-to-light ratio $(M/L)_0$, which is illustrated by the orange area in all panels.
On the other hand, Draco and UMi  orbit solutions imply that they are approaching towards their pericenters, i.e., their $T_{\rm rp}$ are positive. However, TDG3, which is less massive than TDG9, losses its gas at 200 kpc before arriving at pericenter and this causes a large expansion of the galaxy, providing large \mldyn\ values. This TDG3 model fits well all properties of UMi and Dra in Fig.~\ref{figsimobs}, if we take into account the uncertainties of the stellar mass-to-light ratios. In the mass range between TDG3 and TDG9 and for pericenters ranging from  40 to 50 kpc, the properties of those four galaxies are well recovered by our modeling.

\subsection*{Sculptor,  Fornax, Leo I, Leo II}
Sculptor, Fornax, Leo I, and Leo II have relatively large pericenters from 70 to 200 kpc  (Table~\ref{tbdwdyn}). They are bright (i.e., massive) and have relatively high central surface brightness ($<24\,{\rm mag}\,{\rm arcsec}^{-2}$) leading to relatively modest \mldyn\ values from 3 to 11. Their progenitor masses could be intermediate between TDG2 and TDG9, a range that is unfortunately not sampled within the first tidal tail. We do not find any model in Table~\ref{tborb} matching {\it simultaneously} the properties of those 4 galaxies, i.e., in the four panels in Fig.~\ref{figsimobs}. This may be also due to the limited initial conditions of TDG or investigated orbit type, such as a fixed eccentricity at 1.8. Fig.~\ref{figsimobs} shows a model of TDG2 with a 100-kpc pericenter. Its properties are marginally affected by ram-pressure. Looking at the evolution tendencies of all models in panels $c$ and $d$, it is possible, but not proven yet that TDGs less massive than TDG2 could fit those four galaxies with the same mechanism of ram-pressure stripping.

\subsection*{Sagittarius}
Although the Sgr orbit suggests that it is not on the first passage, we cannot immediately rule out that it has the same origin as that of the other satellites. Leo I has passed its pericenter 1 Gyr ago. If Sgr came into the MW halo together with Leo I and at relative lower eccentricity (we discussed this possiblity in Sect.~\ref{secorb}), it may have been captured by the MW and had more than one pericenter passage according to its short orbital period of 0.6 Gyr. In Fig.~\ref{figsimobs}, the properties of Sgr are very similar to those of Carina, UMi, Sextans, and Carina in  $\mu_0$, $\sigma_0$ and \mldyn. The only difference is in absolute magnitude which may be simply due to an observational effect. Sgr is almost 4 times closer than the others on average. With the same depth of star-count observations, one should detect many more stars ($\approx$3 mag deeper) in Sgr over a larger area than in other dSphs. Hence a larger integrated absolute magnitude should be found. This observational effect was discussed in a deep photometric study on UMi by \citet[][see their Sect.\ 4.3]{Palma2003} who argued that the absolute magnitude of UMi may have been underestimated by 1 magnitude. Therefore, the high \mldyn\ and low surface brightness of Sgr are understandable (see Fig.~\ref{figsimobs} and Fig.~\ref{figmuml}) assuming the same mechanism as for the other dSphs, since Sgr is strongly tidally disrupted and has lost its gas recently, within 2 Gyr \citep[see, e.g,][]{Layden2000}.

\begin{figure*}
\includegraphics[width=6.cm]{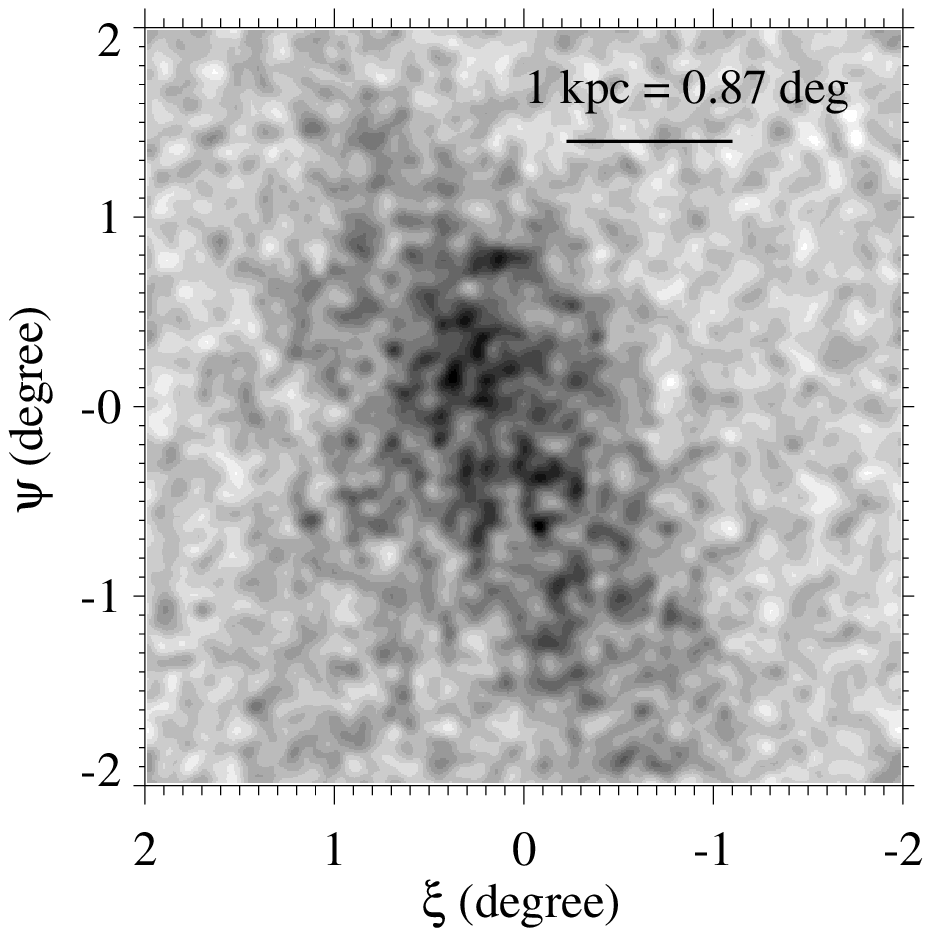} 
\includegraphics[width=6.cm]{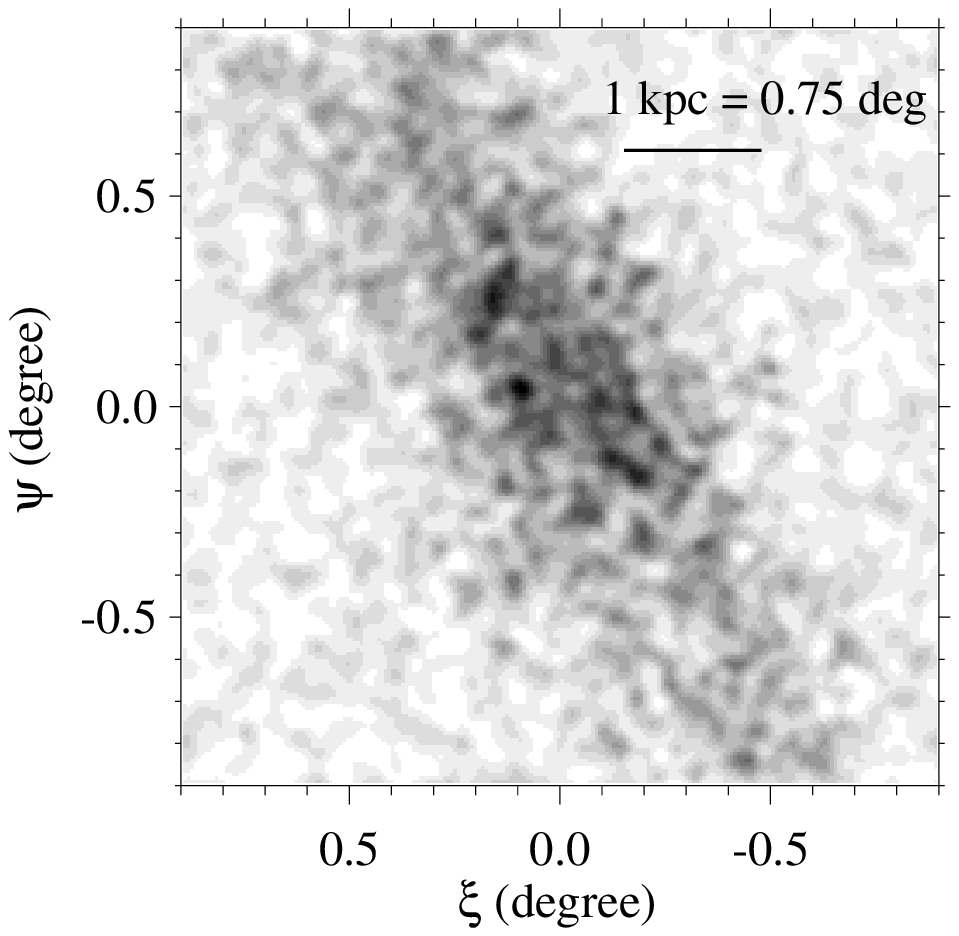} 
\includegraphics[width=5cm]{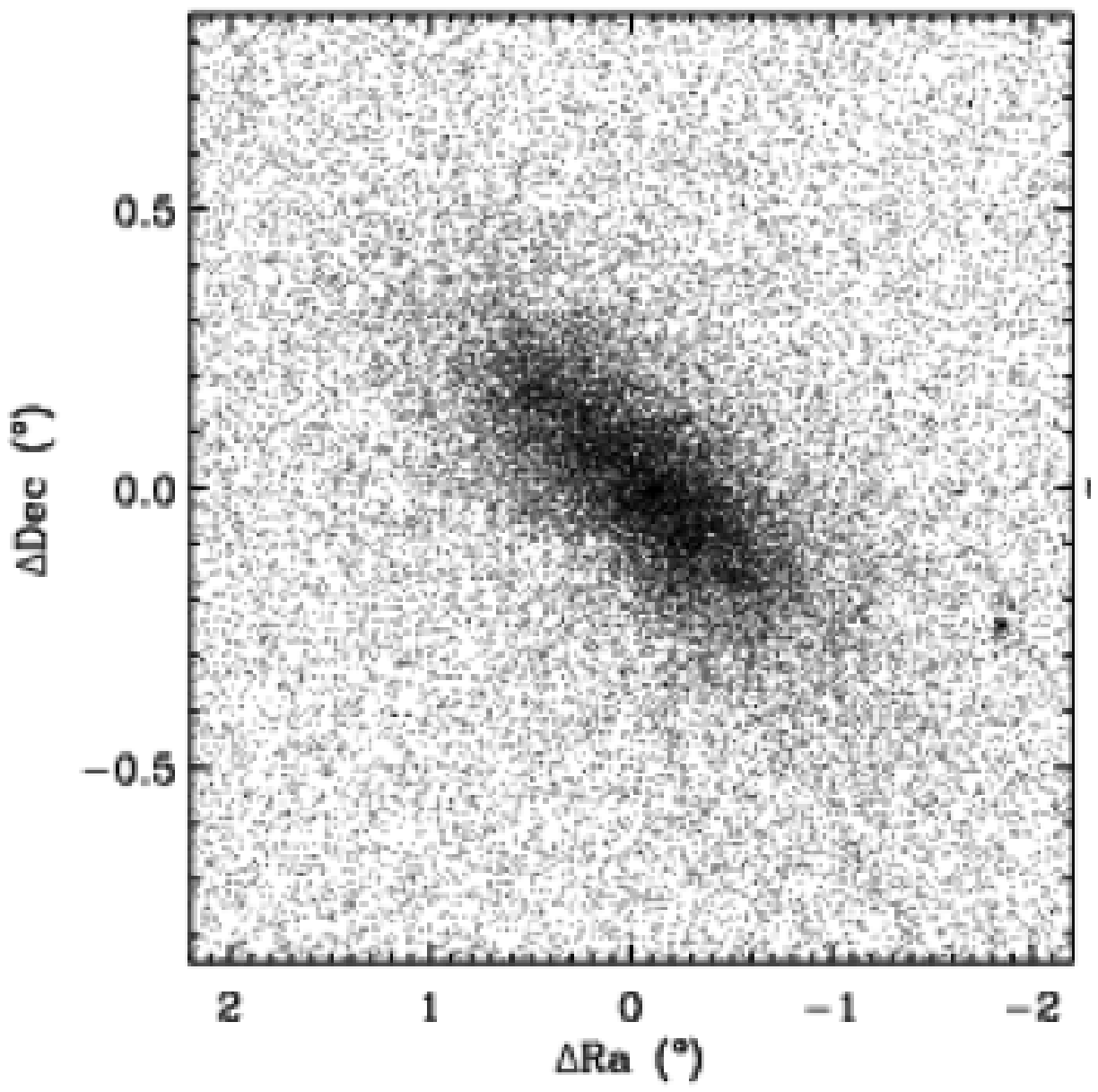} 
\caption{
{\it Left:} Simulated V-band image from TDG9-rp50 when the TDG is located at 68 kpc from the Sun.
{\it Middle:} Simulated V-band image from model TDG3-rp28 when the TDG has a heliocentric distance of 78 kpc. 
{\it Right:} A very deep star-count map of UMi dSph observed at CFHT, 
which is reproduced from \citet{Munoz2012}.  
Note that in the simulated images we added a Poisson noise which corresponds to a detection limits
of 27 (left panel) and $28\,{\rm mag}\,{\rm arcsec}^{-2}$ (middle panel).
}
\label{figimgs}
\end{figure*}

\subsection{Fitting summary and morphological evidence}
With a limited number of TDG ICs and simulations
we tried to reproduce most of the dSph properties, 
including their gas deficiencies, photometries, internal kinematics, orbital parameters,
as well as dynamical mass-to-light ratios.
We reproduced quite successfully UMi, Draco, Carina \& Sextans properties, 
as some of our models directly cover their parameter spaces, simultaneously. 
The mass range of our simulated TDGs is probably too limited to reproduce Sgr, Leo II, Leo I, Fornax, and Sculptor, 
though we identified that they could be explained with the same mechanism with intermediate mass TDGs between TDG2 and TDG9, and possibly with slightly different initial conditions and orbits.
Fig.~\ref{figimgs} shows two simulated V-band images from our models.
They look very similar to some observed MW dSphs, e.g.,
 the image of the middle panel resembles a deep star count map from 
CHFT (Canada France Hawaii Telescope) observation of UMi \citep{Munoz2012}, a galaxy that could be
under strong tidal disruption \citep{2005ApJ...631L.137M}. 
These results suggest that the MW dSphs are possibly the descendants of TDGs, 
just as shown by \citet{1997NewA....2..139K} who have obtained a model with properties nearly
identical to the Hercules satellite discovered 10 years later \citep{2010A&A...523A..32K}.

\section{Discussion}

\subsection{ Capture of TDGs}
From the observations, we have computed the orbits of the MW dSphs (see Table~\ref{tbdworb}). Most of them appear to be on bound orbits, except Leo I and perhaps the LMC, Sextans, and Leo II. Ram pressure induces an internal friction due to gravity between stars and gas when gas is stripped from TDGs, for which we call it as ram-pressure friction. Hence, the initial orbital energy of the TDGs could be significantly reduced, leading to their capture by the MW potential. 
Table~\ref{tborb} shows the setting pericenters (in model names) and compares them to the actual pericenters that are measured from simulations. For TDG9 the actual pericenters are significantly smaller than the setting ones, indicative of a strong dissipation of the orbital energy. 
Interestingly, TDG2 and TDG3 show that their actual pericenters are all larger than the setting ones. This is because TDG2 is too massive to be affected by the ram-pressure friction. On the contrary, TDG3, the least massive TDG, may loose its gas rapidly through ram-pressure stripping leading that the ram-pressure friction acts only for a short time. TDG9 experiences the longest action of the ram-pressure friction (as shown in Fig.~\ref{figsimgas}) leading a strong decay of orbital energy.

Using the same method as in Sect.\ 5.1, we calculated the orbit of TDG9 by taking its position and velocity from the simulation TDG9-rp100 at T=0.5 Gyr after pericenter passage. We find a bound solution with $e=0.85$ while we have input an initial $e=1.8$. This result is confirmed from the simulation itself: 1 Gyr after pericenter passage, TDG9 is fully disrupted and its leading stream falls back to the MW disk, indicating its full capture. Moreover, the orbit solution of Sextans is in agreement with this simulation within one sigma error-bar.
To fully explain all MW dSphs orbital dynamics requires more accurate knowledge of the MW, such as the total mass, the properties of hot halo and the accurate orbits of MW dSphs. The real case could be reasonably complex that some of the dwarf galaxies may have relative motions with respect to LMC, which may change their apparent orbits in the rest frame of MW. SMC might be an example. Nonetheless, our simulations suggest that TDGs may be captured by the MW at the first passage through ram-pressure dissipation, leading to eccentricities comparable to the dSph observations.

\subsection{Revisiting the observational basis of dark matter in dSphs}
The MW dSphs are thought to be dominated by dark matter because of two robust
dynamical properties: (1) large velocity dispersions \citep{1998ARA&A..36..435M} 
and (2) the flatness of radial velocity dispersion profiles 
\citep{2009ApJ...704.1274W,2012arXiv1205.0311W}.
Here, we have shown that these two properties can be explained as well 
by assuming that gas-rich TDGs are progenitors of the MW dSphs (see Sect.~\ref{secres}).

First, \cite{Mayer2001a,Mayer2001b} described well the transformation mechanism from disky dwarfs to dSphs,
and quoting  \cite{2010AdAst2010E..25M},  ``tidal shocks induce strong bar instabilities in otherwise stable, light
disks resembling those of present-day dIrrs. Second, the
bar buckles due to the amplification of vertical bending
modes and turns into a spheroidal component in disks
with relatively high stellar surface density". They also reproduce the overall, regular and elliptical shape of the dSph as it is done in Fig.~\ref{figimgs}, simply because it is the same mechanism. 

Second, the proposed scenario would imply a considerable revision of our understanding of the
Local Group content. Instead of being residuals of primordial galaxies, LG dwarfs would be
mostly associated to TDGs produced during a major collision at the M31 location \citep{Hammer2010}.
\cite{Hammer2013} have shown its uniqueness in interpreting together the M31 Giant Stream, the vast thin disk of satellites around M31 \citep{Ibata2013}, the fact that it points towards the MW,
 the vast polar structure around the MW \citep{2012MNRAS.423.1109P}, the fact that the two vast structures are both rotating, and finally, the proximity of MCs to the MW. 
 The strength of this proposition is coming from the fact that it is simply based on angular momentum of a past collision, i.e. a single origin of
 all the orientation angles linked to the vast relics mentioned above. 
It could be easily falsified given the huge amount of knowledge for the LG content.

Thus, the observational basis of dark matter in dSphs can be disputed by another mechanism that is indeed consistent with our knowledge of M31 and within the hierarchical scenario. 
Purposely, M31 has quite similar fundamental properties (radius, velocity and mass) as most other spirals \citep{Hammer2007}, and a significant part of them were in a merger process 6 billion years ago \citep{Hammer2009,Puech2012}. The M31 bulge, its red outskirts and the large number of tidal streams in its halo are clues to suggest a major event in M31, whose timing is deduced from the stellar ages in M31 halo substructures. 

Evidence is thus required to support the existence of dark matter in the faint dwarfs. This may be alternatively, for example, a positive result for dark matter annihilation towards them, or alternatively a definitive falsification of the M31 merger scenario and the other scenarios \citep[see, e.g.,][]{1997NewA....2..139K,Pawlowski2011}.

\begin{figure*}
\includegraphics[width=7.4cm]{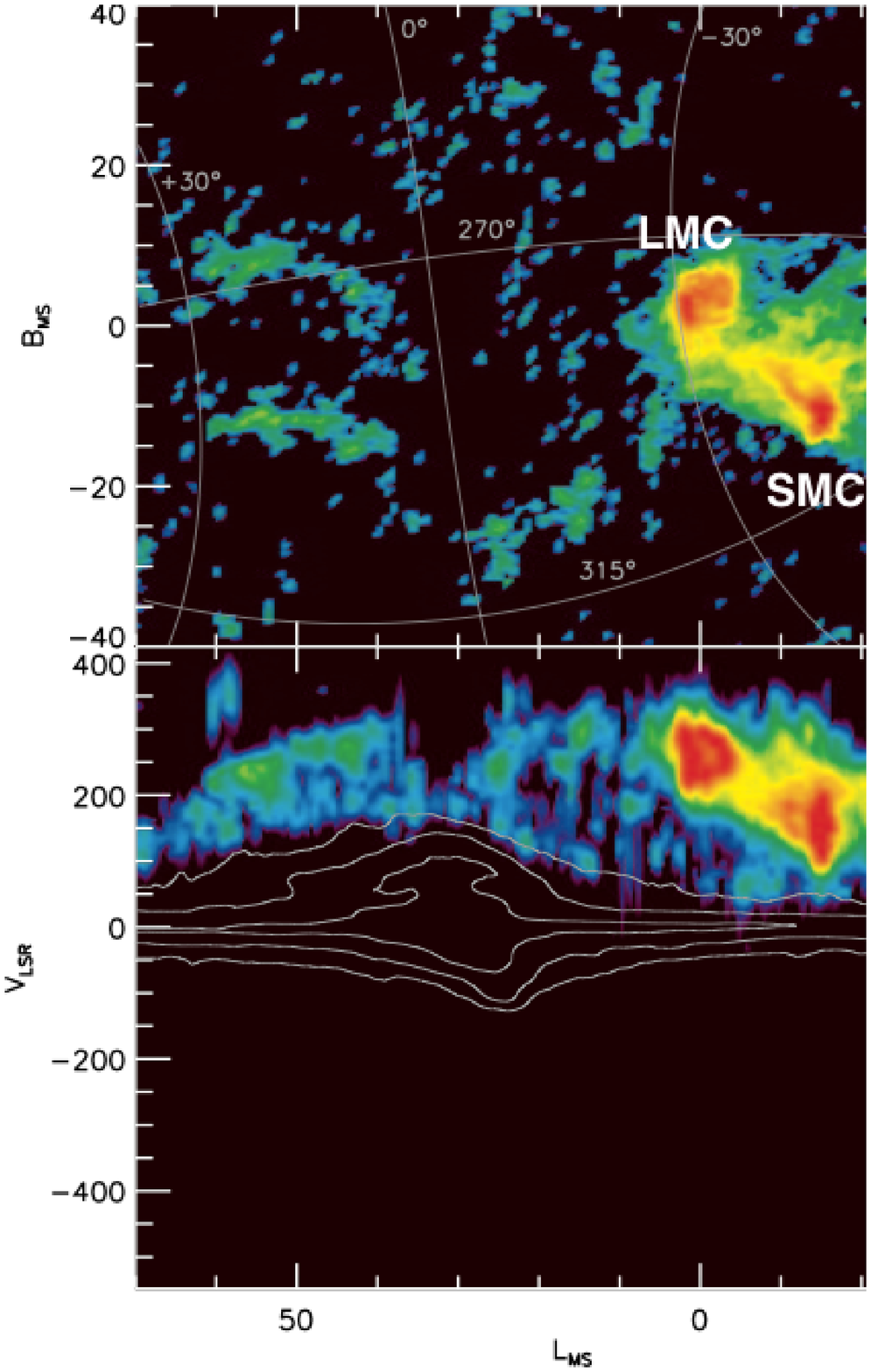}  \ \ 
\includegraphics[width=8.8cm]{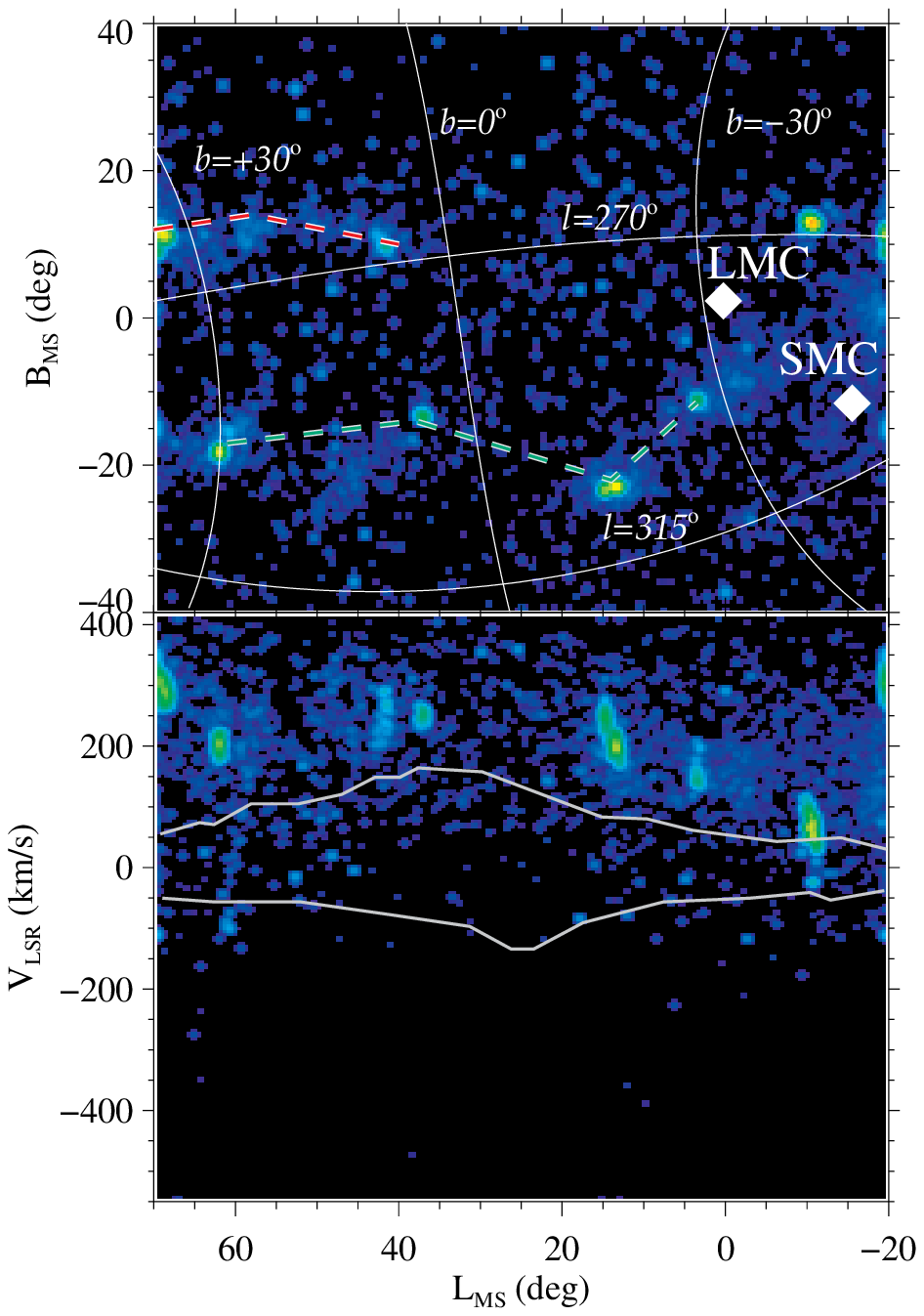} 
\caption{Stripped gas from TDGs, 
as a possible explanation for the Leading Arm (LA) of the Magellanic Stream \citep{Nidever2008,Nidever2010}.
The {\it Left} panels are reproduced from \citet{Nidever2010} figure 8, i.e, the integrated H\,{\sc i} column density map on the sky (top panel) and in L$_{\rm MS}$-$V_{\rm LSR}$ plane (bottom panel) with the MW shown in white contours. Both are displayed in the Magellanic coordinates: longitude L$_{\rm MS}$, latitude B$_{\rm MS}$, and velocity in the local standard rest frame $V_{\rm LSR}$ \citep{Nidever2008}. 
{\it Right} panels are exactly the same figures, but produced by stacking of five simulations. The top panel shows the H\,{\sc i} column density map of the stripped gas from the TDGs, where the positions of LMC and SMC are indicated with white diamonds. The red and green lines connect the clouds that are from a snapshot of a single simulation, respectively. The bottom panel shows a column density map in the L$_{\rm MS}$-$V_{\rm LSR}$ plane from the simulations, where we superpose the outer envelope of the MW form the bottom-left panel. 
}
\label{figLA}
\end{figure*}

\subsection{An observational test for dSph dark matter content and for the missing satellite problem}
From our simulations we notice a feature that could be observationally verified.
After a gas-rich TDG has lost its gas, a destructive expansion process begins.
If dSphs are stable, as assumed in the dark-matter scenario, they will not show any sign of
expansion. If a dSph is expanding, we could detect that the proper motions of stars are proportional to their radius. We tested this idea with model TDG9-rp50 when TDG9 passed the pericenter and arrived at a distance of 95 kpc from the MW. 
We investigated the predicted projected velocities, which could be converted into proper motions, against radius by randomly sampling 300 stellar particles. We found that a mean proper motion may reach $42~\mu{\rm as}$~yr$^{-1}$ at a radius of 2 kpc (1.1 degree on the sky) and a positive correlation between the proper motion and radius could be detected at a 3.6-sigma level. If this expansion really exists in the MW dSphs, such as for UMi which is located 66 kpc, it should be detected by GAIA in the near future \citep{2012Ap&SS.341...31D}. 

As summarized in \citet{Hammer2013}, the scenario of a M31 major merger 
could reasonably well explain several exceptional featured in the Local Group:
the proximity of LMC \& SMC to the MW, the VPOS around the MW, and the recent discovery of VTDS in M31.
\citet{Hammer2013} suggest that the dwarf galaxies in the VTDS could be associated with the loops
formed during the M31 merger.
In the current study, we demonstrate that gas-rich TDGs interactions with the MW
could explain almost all the observational properties of the MW dSphs (Sect.~\ref{secfit}). 
The above suggests that dwarf galaxies around the MW and M31 might be 
TDGs devoid of dark matter, and that MW satellites may be alien TDGs.
It would severely strengthens the existing problem of missing satellites in $\Lambda$-CDM
\citep[see][and references therein]{Klypin1999,Kroupa2012}.

\subsection{The Leading Arm of the Magellanic Stream}
Fig.~\ref{figmor} describes the gas stripping process for a TDG. 
We notice a particularly interesting result that the stripped gas 
may significantly drift from the orbit of TDG 
and tends to be diluted in the surroundings of the MW disk. 
In our M31 scenario, all the progenitors of MW dSphs follow roughly the orbit of the LMC,
and pass through the field of view of the Magellanic Leading Arm \citep[LA,][]{Nidever2008,Nidever2010}.
We thus suspect that this stripped  gas could be linked to the formation of the LA according to their similar distribution features (Fig.~\ref{figLA}). 
The LA is composed of a large number of fragmentary H$\,${\sc i} high velocity clouds (HVCs)
that are broadly distributed over roughly $70\times80$ square degrees on the sky, while its 
velocity distribution is well organized. Actually the features possibly extend to an even larger area (Hammer et al.\ 2013b, submitted).

In the right panels of Fig.~\ref{figLA}, we show a column density map of the stripped gas from the stack of five simulations.
Pericenters of these simulations range from 25 to 90 kpc, in agreement with the values inferred from observations, see Table~\ref{tbdworb}. The orbits of the TDGs in our simulations have the same angular momentum direction (see Sect.\ref{secorb}) while the orbital angular momenta of the MW dSphs are known to point to a similar direction with a dispersion about 30 degree \citep{2012MNRAS.423.1109P,Fouquet2012}. Therefore, in order to produce the best simulated LA, we rotated TDG orbits within a reasonable range, which changes the projected position of the clouds on the sky. We are allowed to rotate those selected simulations because the stripped gas only drifts from the stellar orbits but not strongly interacts with the MW disk yet. As shown in Fig.~\ref{figLA}, the bright and widely distributed features of the LA are almost reproduced by the clouds located in the stripped tail of the TDGs, including their kinematics. Furthermore, clouds in the simulated LA have column densities of $\approx10^{18-20}$~cm$^{-2}$ that are comparable to the observed ones \citep[see][]{Nidever2010}.
Thus, the formation of the LA is possibly linked to the stripped gas from some progenitors of the MW dSphs.

\section{Conclusion}
The recent discovery of a vast thin disk of satellites (VTDS) around M31 \citep{Ibata2013}, together with the vast polar structure (VPOS) around the MW \citep{2012MNRAS.423.1109P}  and the unexpected proximity of the Magellanic Clouds are challenging our understanding of the assembly history of the Local Group. These features are difficult to be reconciled with current cosmological simulations \citep{2010A&A...523A..32K,Kroupa2012}.  \citet{Hammer2013} shows that all these exceptional features could be well unified by a single event that is the ancient major merger that formed M31. The VTDS is consistent with tidal systems which are dominated by the orbital angular momentum of the ancient merger. The M31 merger  model \citep{Hammer2010} predicts that a tidal tail was ejected about 8.5 Gyr ago toward the MW direction and may have currently reached the MW. During the time of travel, gas-rich TDGs have plenty of time to form. The DoS could be formed by the TDGs that were accreted by the MW from the tidal tail has been shown by \citet{Fouquet2012}. The accreted TDGs may have a similar orbital angular momentum as the DoS and that they match the large  velocity of the LMC at its location. 

In this paper, using hydrodynamical/N-body simulations, we describe the in-fall process of TDGs, i.e., their interaction with the MW. We find that due to the ram-pressure stripping from the MW hot halo, gas-rich TDGs may loose their gas during the first passage, which transform them into gas-deficient objects. Some of them may be captured by the MW due to a ram-pressure induced friction between star and gas when they are decoupling. The properties of these objects are comparable to those of the MW dSphs, including their gas deficiencies, large velocity dispersions, absolute magnitudes, central surface brightnesses, heliocentric distances, and orbital pericenters. The properties of UMi, Draco, Carina and Sextans  are indeed well matched by our limited simulations. 
Moreover, we find that the stripped gas from TDGs may fall towards the MW disk and interact with it. This gas may well explain the formation of the Leading Arm of the Magellanic Stream. Both the large dynamical $M/L$ and the flatness of the velocity dispersion profiles are easily reproduced by TDG-MW interactions. This suggests that the dark-matter content of MW dSphs is not robustly determined, as a simple alternative without dark-matter may potentially reproduce all of their properties. 

Given the consistencies between the internal properties of the relics of the gas-rich TDGs compared to the MW dSphs, and its global consistency with our M31 scenario that explains two exceptional anisotropic vast structures, i.e., the VTDS and VPOS (or DoS) in the Local Group, we may conclude that the MW dSphs are possibly DM-free objects because their progenitors could be gas-rich TDGs which are known to be devoid of dark-matter. 
The problem of missing satellites is severely strengthened by our results not only because the MW dSphs are descendants of TDGs
but also because they are possibly alien TDGs ejected from the ancient merger that formed M31. The ``missing satellites problem'' could be better restated as ``a significant excess of small halos'' predicted by the current standard model of cosmology. 

\section*{Acknowledgments}
This work has been supported by the China--France International
Associated Laboratory ``Origins'' supported by the Chinese Academy of Sciences, the National Astronomical Observatory of China, the Centre National de la Recherche Scientifique and the Observatoire de Paris. Part of the simulations have been carried out
at the High Performance Computing Center at National Astronomical
Observatories, Chinese Academy of Sciences.
This work was granted access to the HPC resources of MesoPSL financed by the Region Ile de France and the project Equip@Meso (reference ANR-10-EQPX-29-01) of the programme Investissements d’Avenir supervised by the Agence Nationale pour la Recherche.

\end{document}